\documentclass[pdflatex,sn-mathphys-num]{sn-jnl}

\usepackage{graphicx}%
\usepackage{multirow}%
\usepackage{amsmath,amssymb,amsfonts}%
\usepackage{amsthm}%
\usepackage{mathrsfs}%
\usepackage[title]{appendix}%
\usepackage{xcolor}%
\usepackage{textcomp}%
\usepackage{manyfoot}%
\usepackage{booktabs}%
\usepackage{algorithm}%
\usepackage{algorithmicx}%
\usepackage{algpseudocode}%
\usepackage{listings}%

\theoremstyle{thmstyleone}%
\theoremstyle{thmstyletwo}%

\theoremstyle{thmstylethree}%

\begin{document}

\title[Article Title]{Accelerating Photonic Integrated Circuit Design: Traditional, ML and Quantum Methods}


\author*[1]{\fnm{Alessandro Daniele} \sur{Genuardi Oquendo}}\email{alessandro.genuardi@multiversecomputing.com}
\author[2]{\fnm{Ali} \sur{Nadir}}\email{ali.nadir.ra@hitachi-hightech.com}

\author[2]{\fnm{Tigers} \sur{Jonuzi}}\email{tigers.jonuzi@vlcphotonics.com}
\author[1, 3]{\fnm{Siddhartha} \sur{Patra}}\email{siddhartha.patra@multiversecomputing.com}
\author[1]{\fnm{Nilotpal} \sur{Sinha}}\email{nilotpal.sinha@multiversecomputing.com}
\author[1,3,5]{\fnm{Román} \sur{Orús}}
\author[4]{\fnm{Sam} \sur{Mugel}}

\affil[1]{\orgname{Multiverse Computing}, \orgaddress{\street{Paseo de Miramón 170}, \city{San Sebastián}, \postcode{20014}, \country{Spain}}}
\affil[2]{\orgname{VLC Photonics}, \orgaddress{\street{Off. 0.74 - UPV, Camí de Vera, s/n, Bd. 9B}, \city{Valencia}, \postcode{46022}, \country{Spain}}}
\affil[3]{\orgname{DIPC (Donostia International Physics Center)}, \orgaddress{\street{Paseo Manuel de Lardizabal 4}, \city{San Sebastián}, \postcode{20018}, \country{Spain}}}
\affil[4]{\orgname{Ikerbasque Foundation for Science}, \orgaddress{\street{Maria Diaz de Haro 3}, \city{Bilbao}, \postcode{48013}, \country{Spain}}}
\affil[5]{\orgname{Multiverse Computing}, \orgaddress{\street{192 Spadina Avenue}, \city{Toronto}, \postcode{509}, \country{Canada}}}



\abstract{Photonic Integrated Circuits (PICs) provide superior speed, bandwidth, and energy efficiency, making them ideal for communication, sensing, and quantum computing applications. Despite their potential, PIC design workflows and integration lag behind those in electronics, calling for groundbreaking advancements. This review outlines the state of PIC design, comparing traditional simulation methods with machine learning approaches that enhance scalability and efficiency. It also explores the promise of quantum algorithms and quantum-inspired methods to address design challenges. 
}




\maketitle

\newpage
\begingroup
\setlength{\parskip}{0pt} 
\renewcommand{\baselinestretch}{0.9} 
\small 
\tableofcontents
\endgroup

\section{Introduction}

The growing demand for faster, more efficient technologies in communication, sensing, and computing has spurred significant advancements in integrated circuits. Among these, Photonic Integrated Circuits (PICs) represent a paradigm shift, leveraging the properties of light to enable high-speed, energy-efficient information processing and transmission. Unlike electrons, photons are massless and can propagate without resistive losses, enabling faster data transmission and greater energy efficiency \cite{rudolph2017}. 

Furthermore, the unique properties of light, such as its phase and polarization, offer new dimensions for data representation and manipulation, paving the way for innovations in quantum computing and beyond. With applications ranging from high-speed internet to advanced sensing, PICs have the potential to surpass electronic circuits in areas like bandwidth and power efficiency, increasingly important in today's hyper-connected resource-depleting world \cite{luccioni2023power}.

Despite their promise, the PIC industry remains in its early stages, with design workflows and optimization techniques still lagging behind those of electronics. Traditional simulation methods, while foundational, struggle to keep pace with the increasing complexity of photonic designs \cite{perspective_silicon_photonics, shekhar2024roadmapping}. Machine learning has recently emerged as a powerful tool, automating design processes and enabling scalability, yet it cannot fully address the challenges posed by the intricate nature of photonic systems.

Quantum and quantum-inspired methods offer a new frontier for PIC design by utilizing principles such as superposition and entanglement to explore vast design spaces with exceptional efficiency. Tensor network-based techniques, in particular, enable scalable and accurate simulations of intricate systems, bridging the gap between computational feasibility and complexity. However, a critical challenge lies in coupling classical photonic circuits—where light propagation, fabrication defects, and loss are key considerations—with quantum circuits defined by gate fidelity, coherence, and entanglement. Addressing this intersection could unlock new capabilities in design accuracy, optimization, and hardware performance, creating a synergistic relationship where advancements in one domain propel progress in the other.

\begin{figure}[h]
    \centering
    \includegraphics[width=0.7\textwidth]{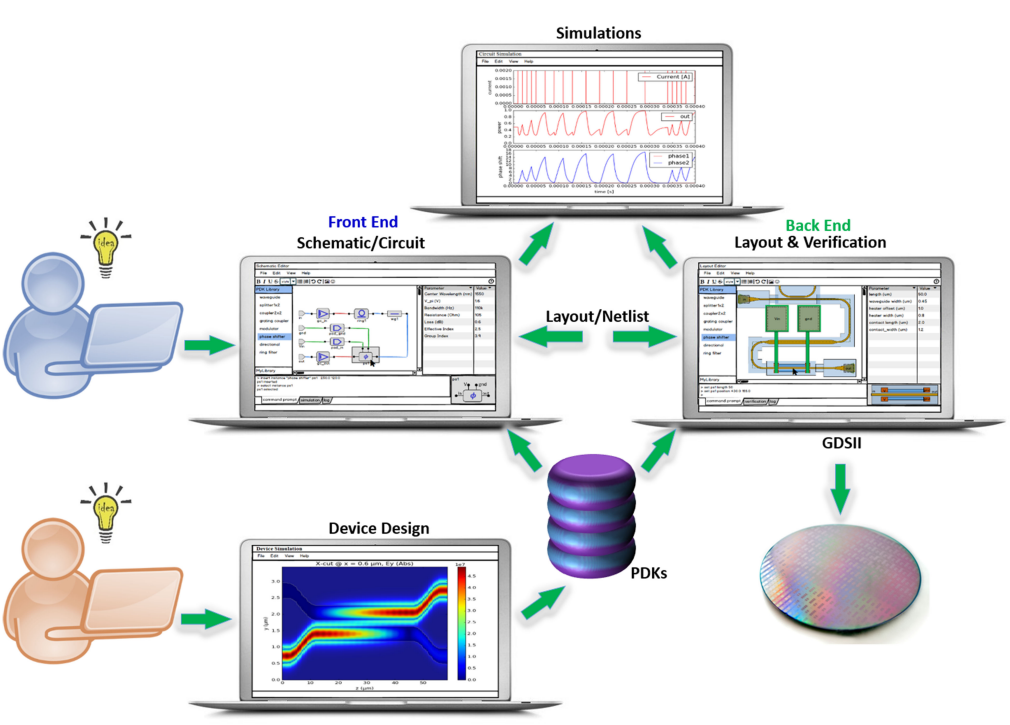}
    \caption{A complete design flow from concept idea to fabrication of photonic circuit. The current photonic design flow uses mostly schematic driven design flow. First, an idea is converted into a schematic. Then, circuit simulations are performed using the compact
models from Process Design Kits (PDKs). The generated circuit is passed to the back end team for layout generation and verification. Fab uses the final mask provided by the back end team for fabrication. At each step photonic tools are used for device design, creation of PDK, and circuit design, simulation analysis and verification.  \cite{khan2019photonic}}
    \label{fig:complete_flow}
\end{figure}

This review begins with an exploration of traditional PIC design methodologies, which have laid the foundation for simulation and optimization workflows in the field. The discussion then transitions to machine learning-driven approaches, detailing optimization algorithms and their applications in photonic circuit design. Finally, the potential of quantum methods is examined, focusing on how these emerging techniques can redefine simulation and optimization processes. Together, these sections provide a comprehensive overview of the current landscape, highlighting the opportunities and challenges for accelerating PIC development and innovation.\label{sec1}

\section{Traditional Design}
\subsection{Device Simulation Techniques}

A cornerstone of photonic device design lies in the ability to predict and model the behavior of light-matter interaction as close to reality as possible. This necessitates advanced simulation techniques that can capture the complexities of electromagnetic fields and their interactions with various materials and structures. Below, we explore the most prevalent simulation methods employed in the field of photonic device design and optimization, each with its unique advantages and applications. We also need to note not only high-performance commercial tools are available, open-access quality simulation tools, using higher-level programming languages such as Python, are also on the rise, widening accessibility or opening new angles to provide more optimal solutions  \cite{laporte2019sr, laporte2019sum}. A comprehensive list of photonic simulation software and libraries is provided in Appendix B. The main simulation techniques used in photonic design modelling are as follows:
\begin{itemize}
    \item Beam Propagation Method (BPM) 
    \item Finite Difference Time Domain (FDTD)
    \item Eigenmode Expansion Methods (EME) 
\end{itemize}

\subsubsection{Beam Propagation Method (BPM)}

The beam propagation method is one of the most popular methods for design and modeling of photonic devices. The BPM method is based on the concept that allows the propagation of an initial electromagnetic field distribution along the propagation direction. The BPM calculates the field profile knowing the refractive index profile of the device \cite{lu2006some}. 

The BPM leverages the fact that waveguide-based photonic devices exhibit a preferred optical propagation direction known as the paraxial direction. This direction corresponds to the primary direction in which optical power is transported, with minimal power flow in perpendicular directions. It is assumed that the refractive index of the material changes gradually along the propagation path in paraxial propagation. This assumption allows for the longitudinal component of the propagating electromagnetic field to be decoupled from the transverse components. 

Therefore, when considering the propagation of the optical field in the longitudinal direction within these optical structures, it is convenient to use the slowly varying envelope approximation (SVEA). Beam propagation methods begin by defining mathematical equations governing the propagation of the electromagnetic field under the SVEA. This approximation neglects the second derivatives of the propagation function of x and y coordinates in the z-direction. The physical meaning of this approximation can be understood by optical field that oscillates rapidly as a function of direction. Its envelope amplitude fluctuates gradually over a period of numerous wavelengths. The main benefit of using slowly varying field instead of electric field amplitudes arises from the fact that factorizing the rapid phase variation enables the slowly field to be represented numerically longitudinal grid that can be much coarser than the wavelength for diverse problems. This contributes to the efficiency of the BPM. There are two main variants of BPM: FT-BPM (Fourier Transform BPM) and FD-BPM (Finite Difference BPM).

\begin{figure}[h]
     \centering
     \includegraphics[width=0.8\textwidth]{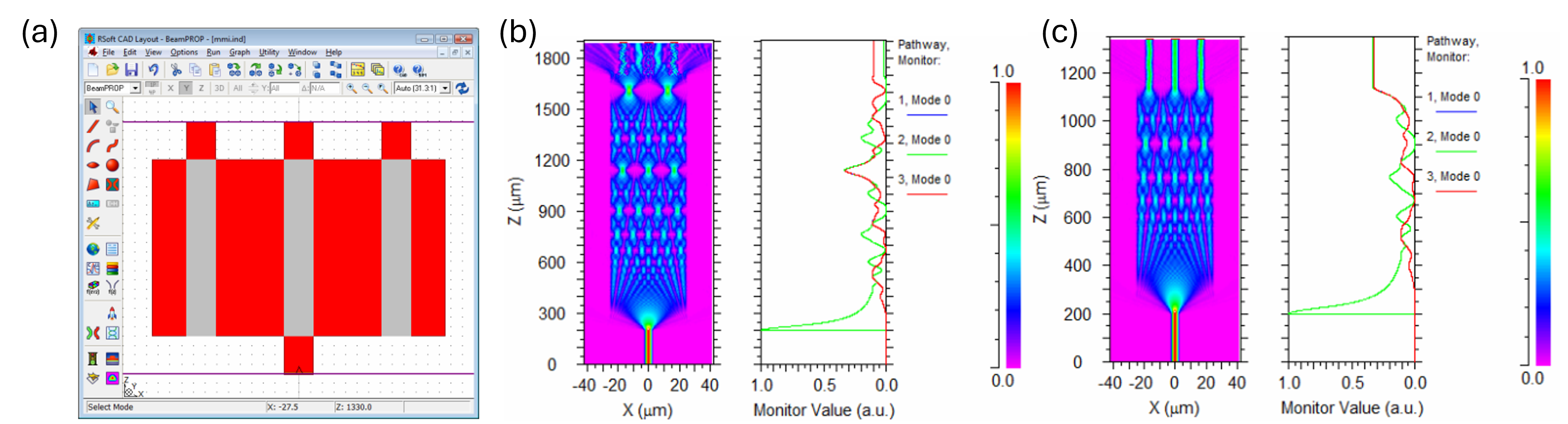}
     \caption{ (a) The 1 $\times$ 3 MMI structure as created in the BeamPROP. (b) The simulation results for the 1 $\times$ 3 MMI device. Note that the best imaging point is where the monitor plots share a maximum value. (c) Simulation results at the optimal length \cite{beamprop}.}
     \label{fig:BPM}
 \end{figure}

Thr BPM method is particularly useful for understanding modal distribution, bending losses, and the effects of waveguide imperfections. Its application is essential in the optimisation of waveguide designs for minimal loss and maximal performance.  The BPM modeling techniques are utilized to design and optimize different parameters of passive photonic devices, such as waveguide, wavelength filters, multimode waveguides, directional couplers and so on \cite{pedrola2015beam}. BeamPROP is commerical design tool based on the Beam Propagation Method (BPM) offered under RSoft photonic device tools. A 1 $\times$ 3 Multi-mode interference (MMI) coupler modeled using BeamPROP is shown in Fig. \ref{fig:BPM} \cite{beamprop}. The MMI couplers leverage a unique integral relationship between the propagation constants of various modes to ensure the self-imaging of the input field along the length of the couplers \cite{soldano1995optical}. The best imaging point, in terms of low loss and minimal imbalance, is achieved when the power output of the three waveguides is nearly the same. The optimal values of $L_mmi$ can then be obtained. The simulation results for the MMI device is shown in Fig.\ref{fig:BPM} (b). Note that the best imaging point is where the monitor have maximum value. The simulation results at the optimal length of MMI is presented in Fig. \ref{fig:BPM} (c). MMI couplers have been implemented to realize a variety of photonic functions, such as splitting, switching and routing. 

\subsubsection{Finite-Difference Time-Domain (FDTD)}

The Finite-Difference Time-Domain (FDTD) method stands out for its versatility and robustness in solving Maxwell’s equations in both time and spatial domains. It is the most widely used numerical technique for simulating propagation of electromagnetic waves in photonic devices. The FDTD scheme was first proposed by Yee \cite{yee1966numerical}. It solves the Maxwell’s equation by discretizing the solution space into spatial grid called the Yee grid. Since its first introduction, the FDTD techniques has seen continuous development. The FDTD employs no potentials and is based on volumetric sampling of the unknown electric fields in the solution domain over a period of time. The solution space is sampled with a resolution finer than the wavelength, as dictated by the design, to accurately capture the highest spatial frequencies of the near field. The simulation time is carefully chosen to maintain numerical stability of the algorithm.

\begin{figure}[h]
     \centering
     \includegraphics[width=0.7\textwidth]{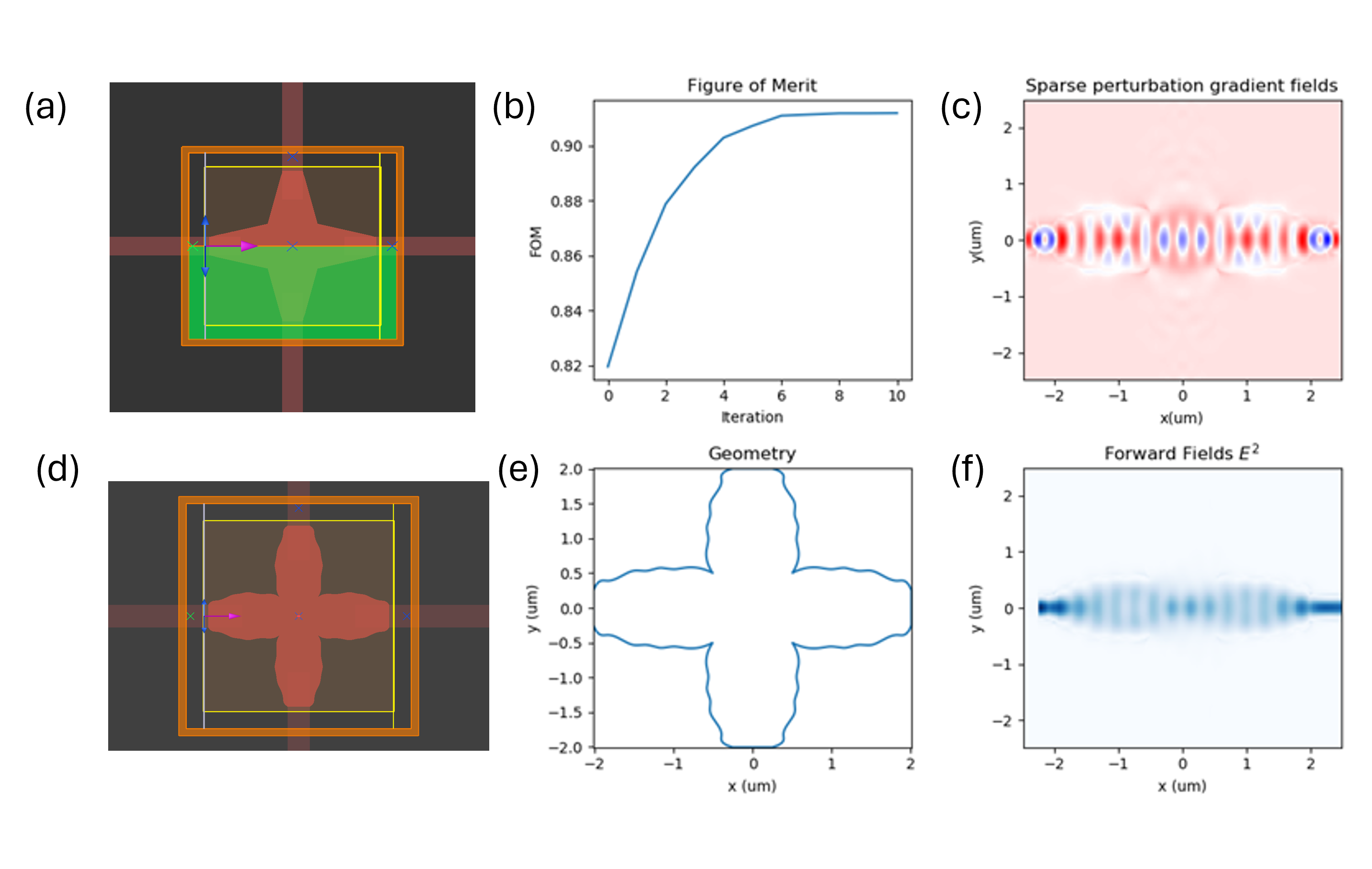}
     \caption{(a) The geometry of the crossing at the start of the optimization process. (b) The figure of merit with number of iteration. (c) The field gradients (due to the changes in geometry) (d) and (e) The final optimized crossing. (e) The forward direction field distributions within the device’s structure obtained using FDTD \cite{fdtd-adj}.}
     \label{fig:fdtd1}
 \end{figure}
 
FDTD provides a comprehensive view of the electromagnetic field behavior within photonic structures, making it invaluable for designing components like photonic crystals, resonators, and complex waveguide systems. The FDTD method can be smoothly combined  with inverse design optimisation techniques to provide unparalleled optimization performance. A waveguide crossing modeled using Ansys Lumerical FDTD and adjoint method optimisation technique is showin in Fig. \ref{fig:fdtd1} \cite{fdtd-adj}. During the optimization process, FDTD simulations provide essential data on the performance metrics of various design iterations, such as transmission efficiency, coupling losses, and field distributions. FDTD results are utilized to calculate the objective function or figure of merit, which is used in the optimization algorithm. This function typically includes parameters like power transmission and insertion loss, guiding the search for the optimal design. This example exhibits the wide applicability of FDTD simulation to a wide range of optimisation tasks and shows their importance in design of high performance photonic devices. There are many reasons for extensive utilization of FDTD and its rapid expansion in photonics device design applications. The FDTD method is known for its accuracy and robustness. The sources of error in FDTD simulations are well characterized and understood allowing for precise modelling across a wide range of electromagnetic wave interaction scenarios.  As a time domain technique, FDTD naturally accounts for the non-linear behavior and directly computes nonlinear response of electromagnetic systems. Its ability to model nonlinear optical phenomena and temporal dynamics makes FDTD a go-to choice for simulations that require a detailed temporal resolution of light-matter interaction. FDTD simulation follows a straightforward systematic approach for device design and optimization. Simulating a new structure using FDTD does not require complex formulation of mathematical equations; instead, the problem is simplified to mesh generation. By using FDTD, the transmission, coupling, resonance behavior, electric field distribution can be easily determined.

\subsubsection{Eigenmode Expansion (EME)}

Eigenmode Expansion (EME) is a simulation technique that excels in the analysis of photonic devices with repeating structures, such as photonic crystal waveguides. EME relies on the decomposition of the electromagnetic fields into a set of local eigenmodes. These eigenmodes exist in the cross section of the device and are calculated by solving the Maxwell’s equations. The modes of a waveguide are consists of few guided modes and radiation modes. EME calculates the eigenmodes of each segment of the device and efficiently simulates light propagation through complex multi-segment circuits by matching these modes at interfaces. Many photonic problems can be fully addressed by considering a small number of modes \cite{gallagher2003eigenmode}. The EME method is fully vectorial and bi-directional. It can take leverage of the symmetry of the geometry to efficiently calculate the waveguides. Fully analytical solutions can be simulated for the modeling of 1D structures. The scattering matrix approach provides a flexible calculation framework, potentially allowing users to selectively recalculate only the modified parts of the structure during the parameter scan optimisation. 

\begin{figure}[h]
     \centering
     \includegraphics[width=0.75\textwidth]{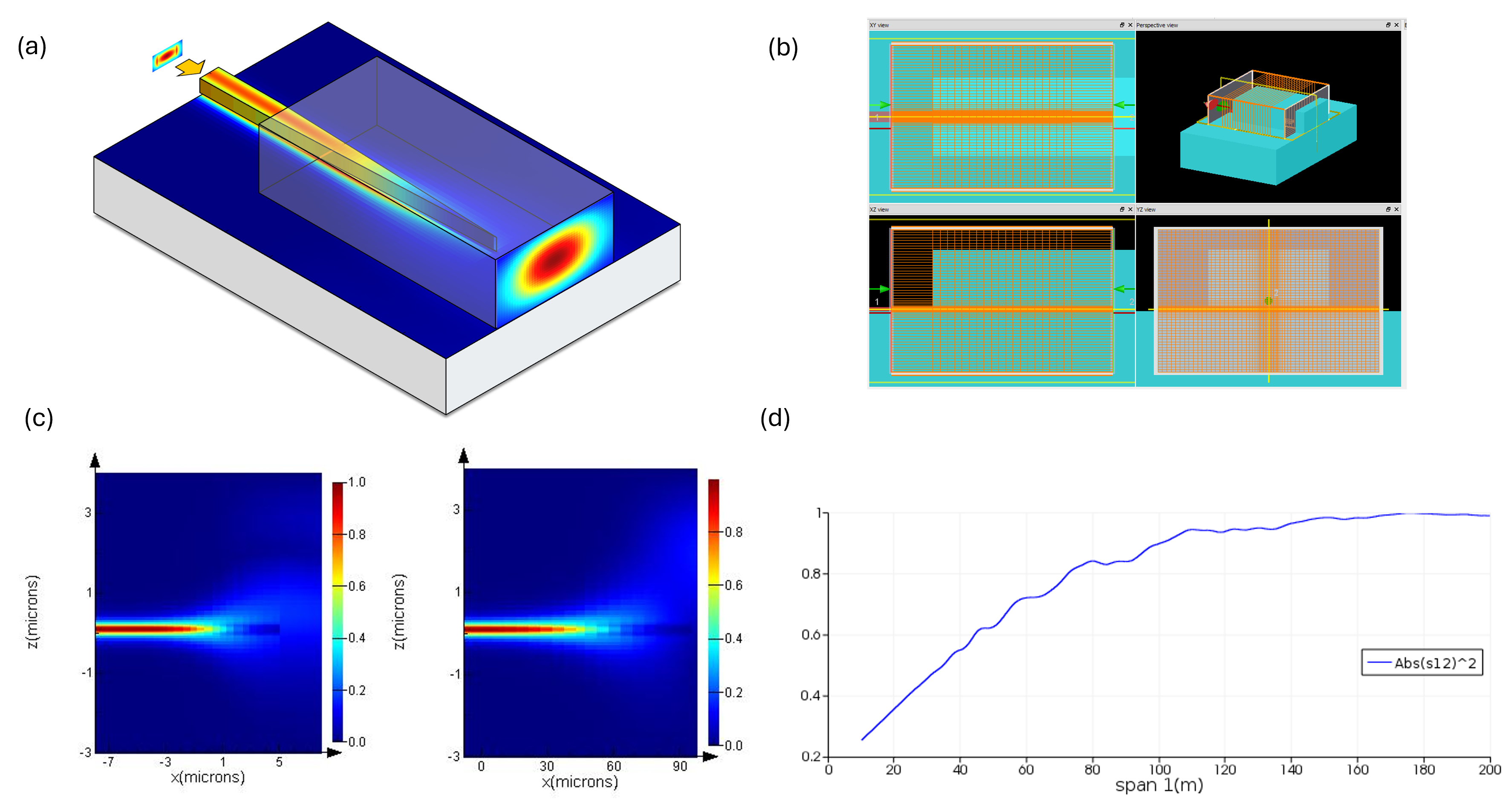}
     \caption{(a) The spot-size converter designed using EME implemented in Ansys. (b) The transverse mesh in the Ansys CAD. (c) The field profile for tapered regions of lengths 10 $\mu$m and 100 $\mu$m. (d) The calculated transmission as a function of varying taper lengths \cite{spot}.}
     \label{fig:EME}
 \end{figure}

A number of photonic devices can be modeled using EME, such as the spot size converter, fiber Bragg grating, MMI Coupler, waveguide Bragg grating, and phase-shifted Bragg grating. An example of a spot size converter designed using EME methods from Ansys is shown in \ref{fig:EME} \cite{spot, tsuchizawa2005microphotonics}. A spot size converter is used to efficiently couple light from a silicon waveguide into an optical fiber. The optical fiber has a much higher mode field size (10 $\mu$m) compared to the silicon waveguide (500 nm). The spot size conversion is realized by using a silicon adiabatic taper covered by a low refractive index waveguide. First, the mode is converted from the silicon waveguide to the low-index waveguide, and then it is coupled much more efficiently into the optical fiber. The EME is best suited for the taper design, as taper lengths can be quickly swept without the need to calculate additional modes. In this case, FDTD-based methods are not computationally efficient because simulation time increases with taper length, and separate simulations will be needed for each taper length. Designing a spot size converter involves several steps. First, a spot size converter is created, which is consist of a substrate, input high index silicon waveguide, the tapered portion of the silicon waveguide, and the low index polymer waveguide as can be seen from the Fig. \ref{fig:EME} (a). In the second step, initiate the EME solver and add monitors to observe the output, and set up ports to launch the light into the structure. Then, the mesh is generated for the entire structure. The mesh can be made finer over the tapered section of the high-index region to better resolve the geometry. Finally,$\mu$mmportant $\mu$esults like S-parameters with elements S11, S12, S21, and S22, as well as the field profile of the device, can be obtained. The parameter sweep of propagation lengths can be run without having to recalculate any modes. The field profile for the 10 $\mu$m and 100 $\mu$m cases is presented in Fig. \ref{fig:EME} (c). To compare the results of EME and FDTD, the EME takes about 3 minutes to simulate 101 different taper lengths, whereas 3D FDTD takes 6 hours to simulate 11 different taper lengths. This clearly shows the superior computational efficiency of the EME method over FDTD for this type of analysis. EME is much more efficient method because it does not require fine discretisation of the solution domain along the direction of propagation \cite{modeeme}. Moreover, the EME method provides a rigorous solution to Maxwell's equations as compared to the BPM, which is only valid under the slowly varying envelope approximation. This method is particularly effective for studying devices that exploit the phase and amplitude of light, offering high accuracy in modeling resonant phenomena and modal interactions. 

\subsubsection{Comparison and Complementarity}

Each of these simulation methods brings distinct advantages to the table. BPM is suited for straightforward waveguide analysis, offering quick insights into light propagation. FDTD provides a more comprehensive approach, capable of capturing dynamic and nonlinear effects with high accuracy. EME, with its focus on modal analysis, is ideal for devices where resonant and phase-sensitive properties are critical. In practice, the choice of simulation method depends on the specific requirements of the design task, and often, a combination of these methods is employed to fully capture the intricacies of photonic circuits. For example, while today there several powerful photonic circuit simulators, not all support more than a single carrier wavelength per simulation \cite{bogaerts2018silicon}. These tools have their strengths and weaknesses, and it can be concluded that none of the discussed tools are universally perfect for all photonic applications. The table \ref{table:1} highlights the strengths and weaknesses of each method and gives an idea for selecting a simulation tool for a specific application \cite{gallagher2008photonic}.

\begin{table}[h!]
\centering    
\begin{tabular}{ |p{3cm}|p{4cm}|p{4cm}|p{4cm}|}
\hline
 & BPM & EME & FDTD \\
\hline
Speed & FD-BPM scales linearly with area & Scales poorly with cross section area & Scales as device volume  \\
\hline
Memory & Usage scales linearly with cross-section area & Increase at the rate between $A^2$ and $A^3$ & Scales as device volume \\
\hline
Numerical aperture & Low NA   & Wide angle beams with high modes & Agnostic to light direction   \\
\hline
$\Delta$n & Better for low $\Delta$n & Accurate for high $\Delta$n & Rigorous Maxwell solver\\
\hline
Polarisation  & V-BPM & Agnostic & Agnostic \\
\hline
Lossy materials & Can model modest losses & Depend on mode solver & Yes \\
\hline
Reflections & Low speed/stability & Easy and stable & Easy and stable   \\
\hline
Non-lenearity & FD-BPM & Difficult to iterate & Easy and stable  \\
\hline
Dispersive materials &  Easy & Easy & Slow down over wide $\lambda$  \\
\hline
Arbitary geometries & Easy & Depend on mode solver & Easy   \\
\hline
ABCs & PML & Depend on mode solver & Yes  \\

\hline
\end{tabular}
\caption{Comparison of simulation methods used for photonic device modelling.}
\label{table:1}
\end{table}




\subsection{ Photonic Circuit Design}
The advanced simulation techniques are instrumental in pushing the boundaries of photonic circuit design. By accurately modeling the behavior of light within these circuits, they enable the precise engineering of devices that leverage the full spectrum of optical phenomena, bringing us closer to realizing the vast potential of photonic technologies. The journey from a mere concept of a photonic circuit to its physical realization and subsequent optimisation is known as design flow. This multi-stage process is fundamental to the successful implementation of photonic circuits, each step bringing the concept closer to reality while addressing the unique challenges posed by photons. Many photonic circuit designers start directly with the layout design of the PIC and only draw circuits to simulate the PIC. This is possible because PICs have components ranging from a few tens to hundreds. This design flow process is similar to the earlier days of electronic IC development.
\begin{figure}[h]
    \centering
    \includegraphics[width=0.7\textwidth]{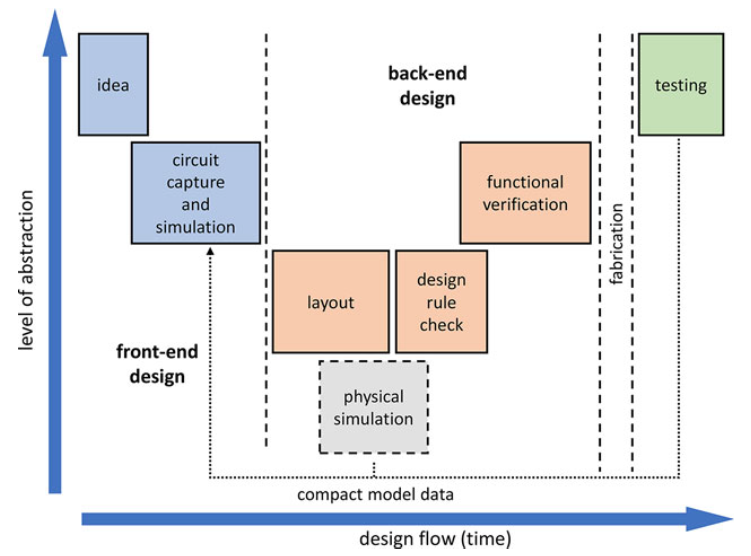}
    \caption{Different levels of abstraction in a photonic circuit design flow. The horizontal axis indicates the sequence of design steps, while the vertical axis indicates the level of abstraction at each step  \cite{bogaerts2018silicon}.}
    \label{fig:design_flow}
\end{figure}
However, in recent past, the complexity and number of components on a PIC is increasing rapidly and necessity the use of design automation to remove the sources of errors. The standard design flow known as the electronic-photonic design automation (EPDA) is necessary to scale PICs and achieve large scale integration similar to the electronic ICs. The EPDA tools follow a specific sequence: first, a schematic is created, and using this, automated processes for circuit simulation, layout generation, layout versus schematic analysis, and design rule checks are carried out. A complete design flow is shown in Fig. \ref{fig:design_flow} and each sections is described in detail in the subsequent sections. 

\subsubsection{Schematic Design}

The conception of a photonic circuit begins with conceptualization, where the foundational idea is defined along with its intended functionalities and performance (design criteria). Following this, a schematic design is drafted, which maps out the circuit in abstract terms, identifying key components such as waveguides, modulators, detectors, and their interconnections. This stage is critical for setting the framework within which the circuit will operate, akin to the blueprint of a building before construction begins.  The idea is to connect known devices in such a way that some target functionality is achieved. The current photonic integrated circuits have tens to thousands of components in a single chip. The circuit size is increasing with technology platforms like silicon photonics. With advancements in photonics technology, the complexity of photonic circuits is increasing, which is closely linked to the design tools that assist designers in reliably expanding their circuit designs. Therefore, photonic tools should evolve to accommodate more complex functionalities, enabling optimise performance and innovate solutions that meet the demands of future applications.

\subsubsection{Circuit Simulation} 

Simulation acts as the bridge between schematic design and physical realisation. Given the complexities of light-matter interactions, photonic circuits require precise and accurate simulations to predict how they will perform and predict points of failure.  These simulations are vital for refining the design, ensuring that the circuit will function as intended once fabricated. Unlike the device simulation, the circuit simulation is carried out by abstract behavioral response or compact models, which maps outputs to inputs. The compact models are offered by the foundries and their quality dictates the effectiveness of the various circuit simulation tools. As of today, both commercial and open-source circuit simulation tools are available \cite{lumerical, vpiphotonics, cadence, gdsfactory, optiwave, picwave, synopsys}.

\subsubsection{Process Design Kit (PDK)} 

\begin{figure}[h]
    \centering
    \includegraphics[width=0.7\textwidth]{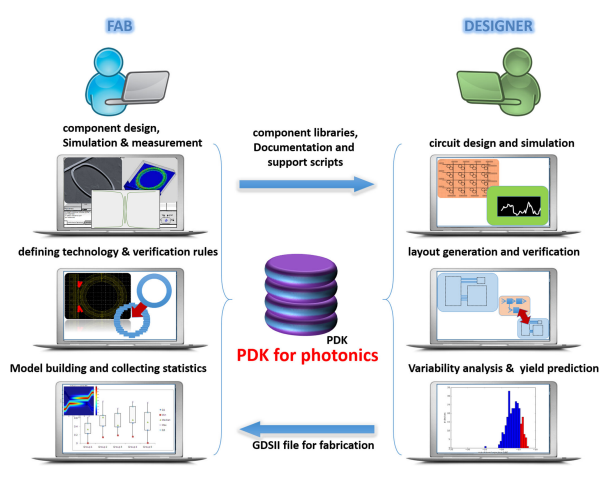}
    \caption{ The PDK contains the necessary information about the devices and manufacturing process offered by the foundries \cite{khan2019photonic}.}
    \label{fig:PDK}
\end{figure}

PDKs contain the necessary information to design circuits, including device models, building blocks, design rules, verification deck, process parameters, etc. The PDKs are essential for photonic circuit manufacturing as they act as a bridge between the foundries and the designers, enabling the reliable fabrication of circuits. Designers use PDKs throughout the design process, from schematic design to circuit layout and verification as represented in Fig \ref{fig:PDK}. It allows foundries to include the essential properties of technology platform without revealing the intellectual property to the customers. In some fabs, the building blocks are represented as black box and behavioral models are compiled in such a way that the important parameters and equations are not accessible to the circuit designer.

\subsubsection{Circuit Layout}

With a refined schematic bolstered by simulation data, the layout design process turns abstract concepts into a tangible form for manufacturing. This involves detailed planning of optical paths, component placement, and minimizing losses and cross-talk - all within the constraints of available fabrication technologies. Waveguides need to maintain a minimum distance to prevent cross-talk. Some photonic circuit tools automatically route with different curve geometry, however, most waveguide routing is done manually \cite{bogaerts2016ipkiss, ghillino2018synopsys}. The layout is created using open artwork system interchange standard (OASIS) or graphic data system (GDSII) format \cite{klayout}.

\subsubsection{Circuit Verification}

All steps command attention to detail and reduce inefficiencies to a minimum, ensuring that their intrinsic advantages, speed, bandwidth, and efficiency are fully realised. The generated layout is then checked for potential rule violations and errors of the design rules supplied by the foundry. The design rules typically include specifications for the minimum critical dimensions, acceptable sharp angles and layer density requirements. These design checks are performed by the design rule checking software tools provided by Cadence Physical verification system \cite{cadence} and Synopsys IC validator \cite{synopsys}. The foundries expect a design with clean DRC and provide design rule checking guidelines to be used with verification tools. Finally, the layout file is sent to the foundry for fabrication. Fabrication brings the PIC into existence, using processes such as lithography, etching, and deposition to construct components and pathways for guiding and manipulating light.

\subsubsection{Testing and Optimisation}

The final stages of the design flow involve testing the fabricated circuit to evaluate its performance against the initial specifications and optimising it to enhance its functionality. This may include tuning the properties of components, adjusting layouts, or even revisiting the schematic design based on the outcomes of real-world testing. The goal is to create a functioning chip by considering the real impacts of the fabrication process. Variability analysis is important to understand how non-idealities in circuits accumulate and affect signal propagation. This is achieved by mapping variablity from device level parameters to circuit level performance and require extensive data collection through characterization. To accurately predict circuit performance, models need to account for variablity at different abstraction levels. Monte-Carlo simulation are commonly used as assess the performance of circuit by performing repeated test based on randomised lower level parameters. However, this process can be computationally intensive. Emerging techniques like polynomial chaos expansion and  stochastic collocation techniques have been used to enhance the performance by mapping parameter distribution directly to performance matrics.  Improvements are needed to integrate efficient simulation techniques with methods that account for parameter correlations and spatial dependencies. It is in this phase that the potential of machine learning and quantum computing becomes most apparent, offering advanced tools for analysing performance data and identifying optimal strategies \cite{bogaerts2018silicon}.
\label{sec2}

\section{ML-driven Design}

The traditional photonic device optimisation approaches start from an initially known physics-inspired method, and then specific parameters are optimized for a suitable application by running a parameter sweep of relevant characteristic parameters. The simulation tools are used to solve Maxwell’s equations for the initial designs. The adjustment of some parameters and reevaluation are performed to approach the desired target. This technique has been extensively used, resulting in various novel devices, achieving remarkable results, and producing a large library of templates. However, with increasing demand of better devices with larger bandwidth, smaller footprint, and high integration, continuing with the traditional approach becomes computationally costly, time-inefficient, and increases complexity. The traditional photonic design library lacks techniques that can provide optimal design in which several independent characteristics can be simultaneously optimized.

A different approach known as the inverse design approach handles these tasks in a different manner. This technique does not require any physical principle for the initial guess, the target photonic functionalities are achieved by optimizing in the design parameter space. The desired solution is achieved by using advanced algorithms to provide a solution that maximizes or minimizes a function related to the target values of a parameter. This is a non-intuitive approach that allows to find optimal design by searching in the full parameter space. The use of inverse problem formulation for photonic optimisation was first carried out by Dobson and Cox  \cite{dobson1999maximizing} and Spuhler et al. \cite{spuhler1998very}.  

The most popular algorithms used till date can be divided into two categories: the evolutionary methods and gradient-based methods. Evolutionary methods include genetic algorithms and particle swarm optimisation, while gradient-based methods include topology optimisation, steepest descent, and so on. Some other noted approaches based on nonlinear search or heuristics have also been utilized for photonics. The iterative algorithms become extremely laborious when a single device needs to be optimized for a parameter that slightly differs, for example, grating couplers working at different wavelengths. A deep neural network (DNN)-based tool can eliminate these problems by mapping direction between the geometric parameters and optical response. It eliminates the need for optimisation and can generate a device for a specific target using a well trained DNN model. In the next section, we will summarize the recent advances in photonic device design approached utilizing these optimisation techniques \cite{shen2024recent}.

\subsection{Global Search Algorithm} 
\begin{figure}[h]
     \centering
     \includegraphics[width=0.7\textwidth]{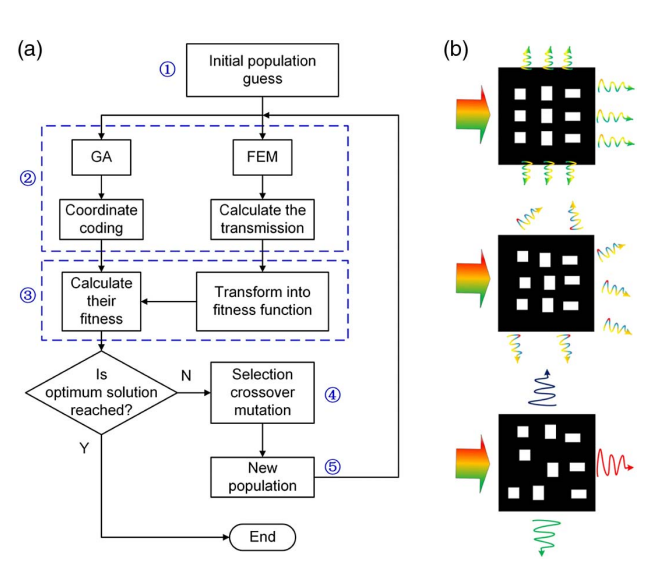}
     \caption{ (a) The optimisation process of GA. (b) Evolution of the device structure at each step of the optimisation process
     \cite{liu2019integrated}.}
     \label{fig:ga_flow}
 \end{figure}
Genetic algorithms (GA) use a searching strategy based on a natural selection process that mimics biological evolution. GA have been widely used to design QR-codes and empirical structures with different ways of encoding the design area \cite{whitley1994genetic}. In the case of QR-code structure, the GA algorithms divide the design area into thousands of pixels of binary array. The generated structure is based on the arrangement of binary pixels designated "0,” which represent a one type of physical space, and "1," which represents another \cite{sanchis2004integrated}. In the past, GA has been extensively used to design and optimize a range of photonic waveguide devices such as waveguide filter \cite{liu2005compact}, sensor \cite{bahrami2013improved}, polarization converter \cite{correia2003genetic}, taper \cite{haakansson2005high}, wavelength router \cite{liu2019integrated}, etc.

\begin{figure}[h]
     \centering
     \includegraphics[width=0.7\textwidth]{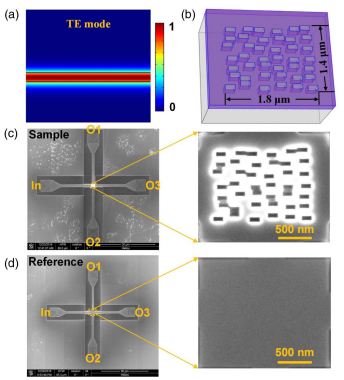}
     \caption{SEM images of the wavelength router designed using GA optimisation \cite{liu2019integrated}.}
     \label{fig:ga_example}
 \end{figure}

A flow chart for the optimisation of broadband wavelength routers using GA and photonic simulations based on the finite element method is shown in Fig. \ref{fig:ga_flow} (a) \cite{liu2019integrated}. The first step in optimisation is to consider a planar binary structure composed of two different materials. The whole structure is divided into basic cells of any arbitrary shape or any defined shape like a circle, square, rectangle, and so on. The size of the basic cells can be controlled considering the fabrication constraints. By changing the cell distribution in the structure, the electromagnetic field distributions can be altered. Thus, the transmittance of the light propagating through the structure can be manipulated to achieve the desired target. The distribution of the cells is described as the optimisation design variable. The channel transmittance is treated as the optimisation target that provides the performance of the device. As shown in the flow chart Fig. \ref{fig:ga_flow} (a), a random initial sample of structures (population) is generated as the first generation in step [1]. These structures are created by considering different models with different irregularly arranged structures containing different arrangements of basic cells.  In step three, the GA is updated by the electromagnetic field response solved by the finite element method and calculates the fitness function of the individual structures. The fitness function output values represent the target result. Then the population is sorted based on the score in step four. The population with higher fitness values is selected as the parent generation. In step five, GA produces a new child generation by creating combinations of parent code through crossover and mutation. Similarly, new structures with better performance are generated. This will be carried out until the specified target performance is achieved. During this loop, the structures underwent changes, as shown in Fig. \ref{fig:ga_flow} (b). An optimized structure working as the nanophotonic router will be obtained in which the incident wavelengths are separated into corresponding channels. The SEM images of the optimized router device are shown in Fig.\ref{fig:ga_example}.

\subsection{Particle Swarm Optimisation} 

Particle swarm optimisation (PSO) is based on the social behavior and interaction of bird flocks or schools of fish \cite{gad2022particle}. The algorithm starts by randomly generating a set of potential solutions, called "particles," within a defined parameter space. The whole parameter space is occupied by a collection of solution, called particles, that sits at random positions with random velocities. Each particle in parameter space represents a possible device structure. A figure of merit defined considering the target specification is used to evaluate the position of particles. The FOM is a mathematical function that quantifies the performance of a particular design based on desired criteria. For instance, in a power splitter, the FOM could be designed to favor high transmission efficiency and low back reflection \cite{zhang2013compact}. This calculation involves running simulations using FDTD or other methods.  Throughout the optimisation process, PSO keeps track of the best position ever achieved by any particle in the swarm, known as the "global best" ($G_{best}$) Simultaneously, each particle remembers its own personal best position ($P_{best}$), representing the parameters that yielded its highest FOM so far. Based on the locations of $G_{best}$ and $P_{best}$, each particle adjusts its velocity and position in the parameter space. The positions and velocities of each particle are calculated using a mathematical equation and updated accordingly. The solution is achieved by gradually approaching the target function optimum by repeating the optimisation process. The algorithm typically terminates when  a maximum number of iterations are reached or achieve a satisfactory FOM. This process is analogous to how birds in a flock adjust their speed and direction based on the positions of the flock leader and their own experiences \cite{gad2022particle}.

\begin{figure}[h]
     \centering
     \includegraphics[width=0.75\textwidth]{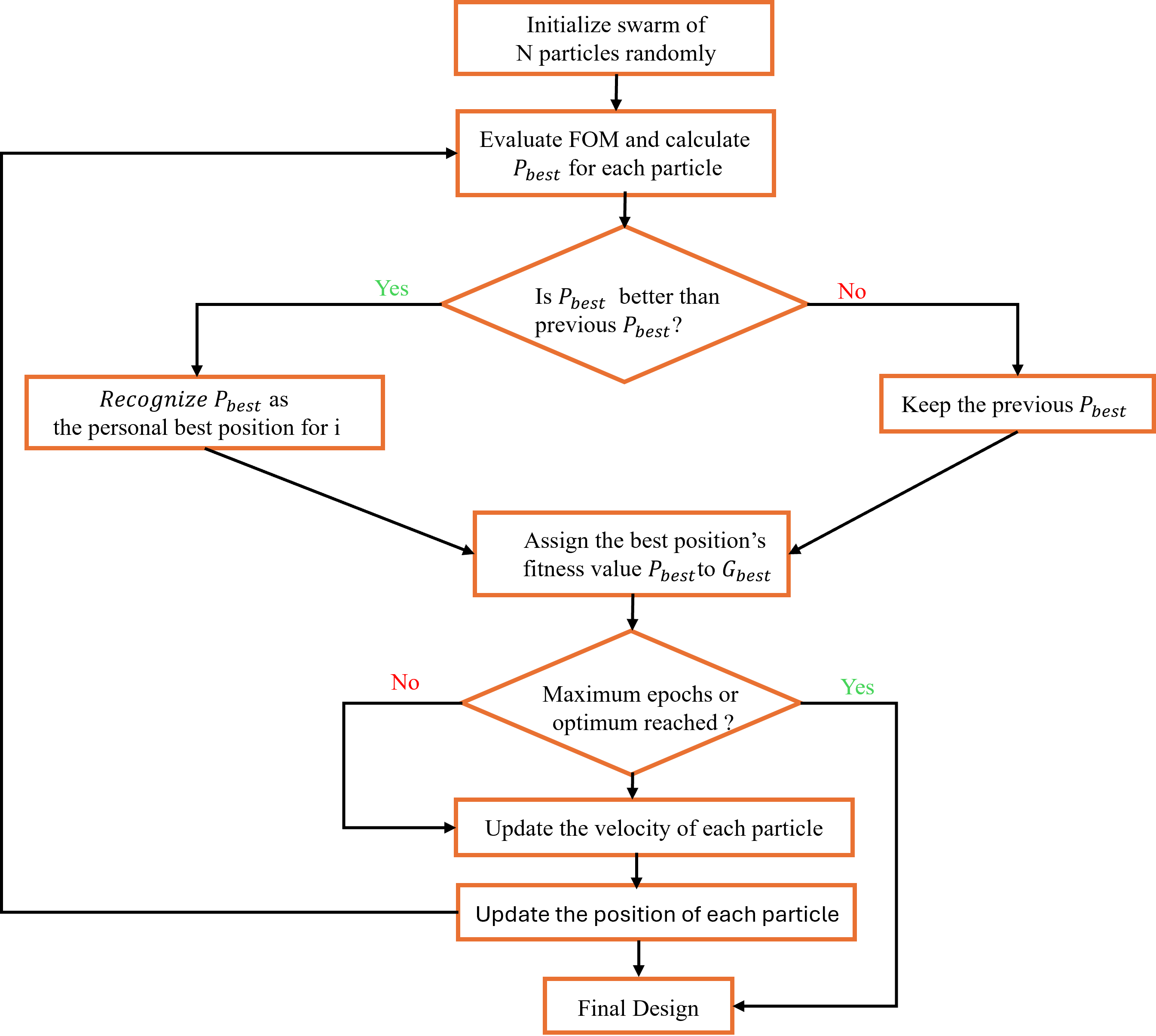}
     \caption{The optimisation process of PSO.}
     \label{fig:pso_flow}
 \end{figure}

Key elements of PSO are the communication between the particles and the capability to learn from the swarm's collective experience. The main advantage of PSO is having fewer parameters to tune, which simplifies the setup process. However, it converges slowly in high-dimensional search spaces and fails to find the global optimum due to local optima traps and fluctuating particle velocities \cite{gad2022particle}. A flow chart of the PSO process is shown in Fig. \ref{fig:pso_flow}.

\begin{figure}[h]
     \centering
     \includegraphics[width=0.8\textwidth]{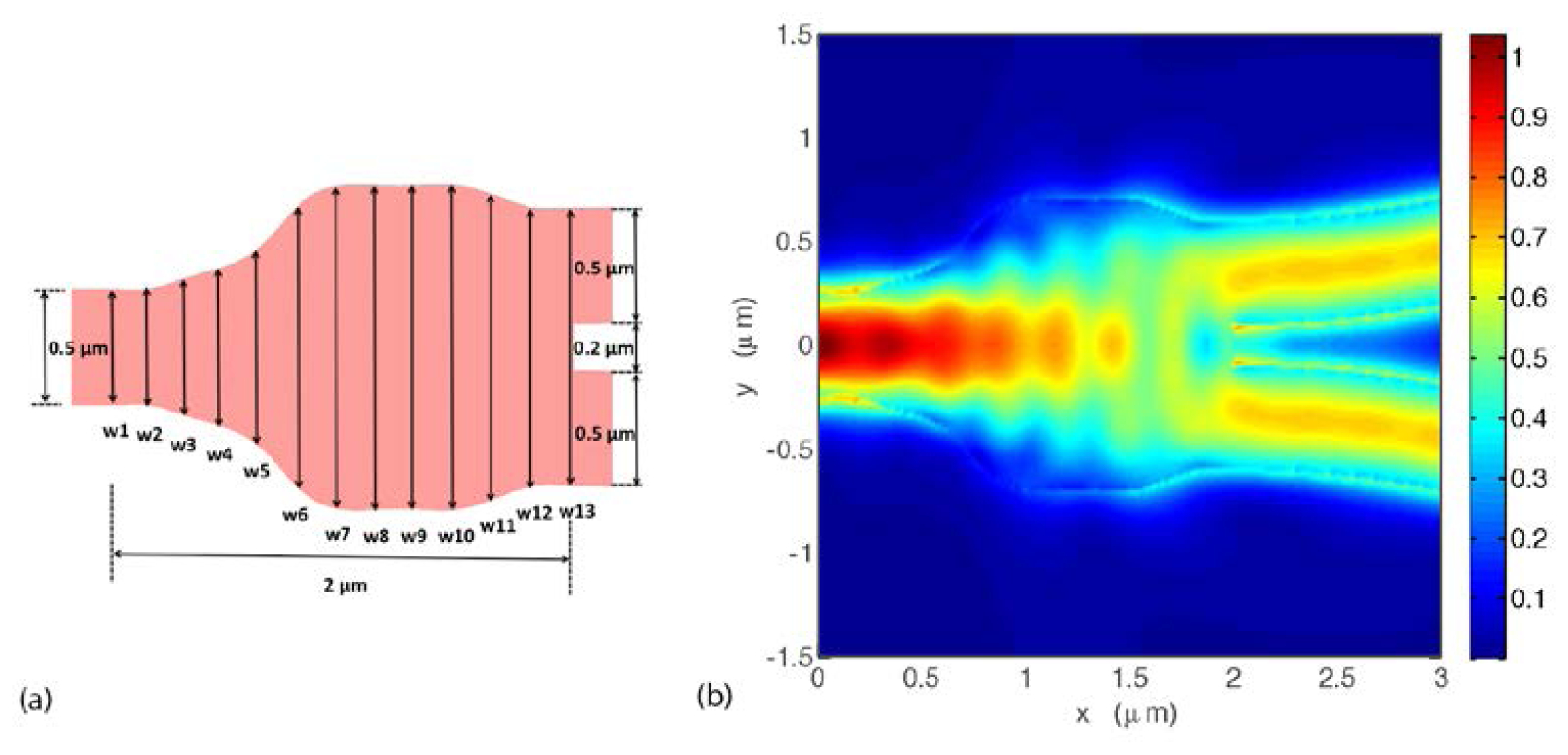}
     \caption{ (a) A Y-junction power splitter divided into segments of equal length for the PSO optimisation. (b) Electromagnetic field simulation of the optimized Y junction \cite{liu2019integrated}.}
     \label{fig:pso_example}
 \end{figure}

 PSO algorithms have been used for the design of custom integrated photonic devices. In this application, each particle represents a photonic system within a parameter space. The merit functions calculate the quality of the photonic system. The effectiveness of the PSO is demonstrated through an assortment of application areas such as wavelength diplexer \cite{ma2014design}, asymmetrical directional coupler \cite{wang2019broadband}, mode-order converters \cite{guo2023experimental, chen2015low}, grating couplers \cite{zhang2022characteristic}, polarization beam splitters \cite{lu2018particle}, and multilayered optical filters \cite{lee2023implementation}.

Using this approach, a Y-junction power splitter for submicron silicon waveguide was optimized by dividing the whole structure into multiple sections of the same length. As shown in Fig. \ref{fig:pso_example} (a), the taper was divided into 13 segments. The width of each section is optimized to achieve low-loss coupling. The optimisation FOM was defined as the power in the $TE_0$ mode at either branch, calculated using the overlap integral between the $TE_0$ mode of the waveguide and the detected field at the output branch. The FDTD method was used for computing efficiency during the process (Fig. \ref{fig:pso_example} (b)). Using the PSO optimisation technique, they achieved a low insertion loss of about 0.28 dB with a device footprint of 1.2  $\mu$m x 2 $\mu$m, an order of magnitude smaller than typical multimode interferometers \cite{zhang2013compact}.

PSO technique is robust to manufacturing errors because fine-tuning of empirical structures does not produce small, difficult-to-fabricate features. PSO only uses one operator, velocity calculation, which reduces computation time . However, it is essential to note that the success of PSO heavily depends on defining a suitable FOM that accurately reflects the target device performance. It requires prior knowledge about the devices to be designed, such as the empirical initial structures and the key parameters \cite{huang2022new}.

\subsection{Direct Binary Search} 

Direct binary search (DBS) is a nonlinear optimisation algorithm used in the inverse design of photonic devices. The DBS algorithm sequentially explores every element of the design space of possible pixels based on a predefined figure of merit. The objective function, or figure of merit, serves as a critical parameter in its design and varies according to distinct optimisation objectives. The FOM is typically defined as a function of the desired optical properties, such as transmission efficiency, crosstalk, or bandwidth. 

\begin{figure}[h]
     \centering
     \includegraphics[width=0.9\textwidth]{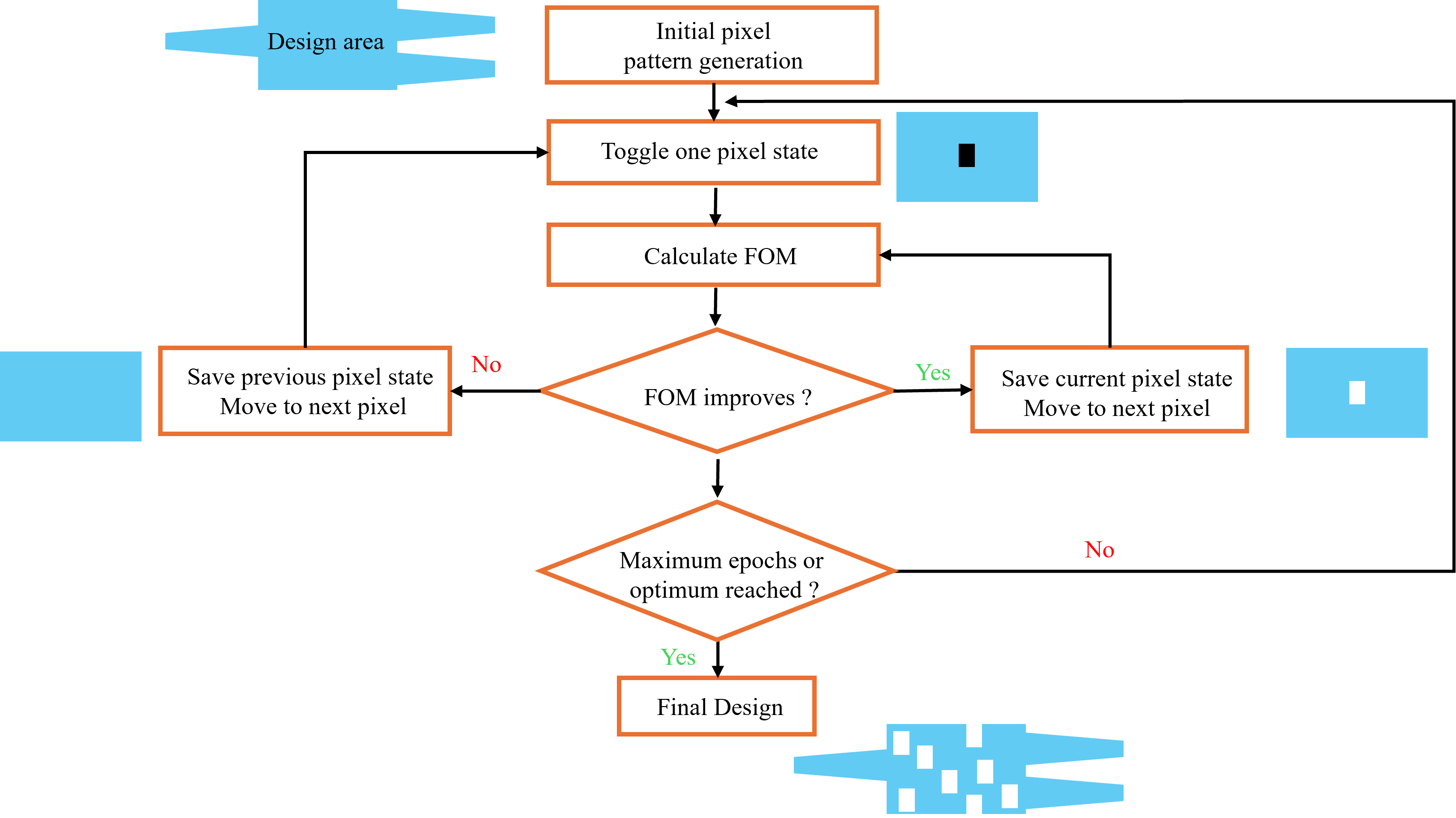}
     \caption{Optimisation process of DBS and changes in the evolution of device structure.}
     \label{fig:dbs_flow}
 \end{figure}
 
First, a random initial pixel distribution matrix is created in the device region.  For instance, a device with dimensions of 3 $\mu$m $\times$ 3$\mu$m could be discretized into a grid of 30 $\times$ 30 pixels of 100nm $\times$ 100nm dimensions.

Each pixel state can represent two states; for example, the “0” state can represent silicon material and the “1” state can represent silica material or air.  During each iteration, the state of a randomly chosen pixel is altered. Then the figure of merit is calculated by altering the state of the pixels using numerical methods like the FDTD or BPM. If this process results in an improved figure of merit, the updated pixel state remains the same; otherwise, it is reverted to its original state. One iteration completes when every pixel has undergone a change in state. These iterations persist until the FOM reaches a predefined threshold or a maximum number of iterations is reached \cite{shen2015integrated} . A process flow of DBS optimisation is presented in the Fig. . A number of photonic devices including power splitters \cite{ma2020inverse, ma2020arbitrary}, crossings \cite{lu2017inverse, chang2018ultracompact, shen2015integrated}, switches \cite{ma2021inverse}, polarization beamsplitter \cite{chang2020inverse}, power splitters \cite{chang2018ultra, ma2023different} have been demonstrated. 

\begin{figure}[h]
     \centering
     \includegraphics[width=0.9\textwidth]{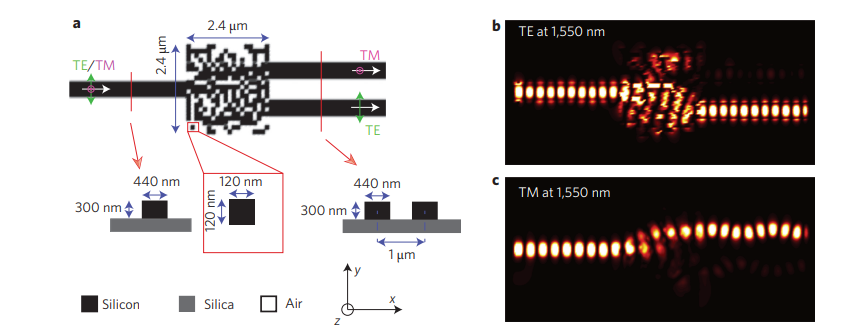}
     \caption{ (a) An integrated-nanophotonics polarization beamsplitte designed using the DBS optimisation. (b) Field simulation showing the seperation of TE and TM mode into two output waveguides of MMI \cite{shen2015integrated}.}
     \label{fig:dbs_example}
 \end{figure}

In reference \cite{shen2015integrated}, a polarization beam splitter (PBS) was designed using the DBS algorithm as shown in Fig. \ref{fig:dbs_example}. The PBS split unpolarized light into its transverse magnetic (TM) and transverse electric (TE) components and guided them into separate waveguides. he optimisation process begins by discretizing the device into pixels. The device is patterned as a 20x20 grid of square pixels with 120nm sides. The FOM is calculated at the transmission efficiency for both TE and TM polarization states. The FDTD was used to simulate the structure at each iteration to calculate the FOM. The desinged device showed transmission efficiencies of 71\% and 80\% for TE and TM modes, respectively. The corresponding calculated extinction ratios were 11.8 dB and 11.1 dB, for the TE and TM modes, respectively.

One of the main advantages of DBS is that it can be used to design devices with very compact footprints. DBS enhances the design figure of merit significantly, scaling from a few parameters to hundreds of pixels, thereby expanding design flexibility and enhancing device performance. 

Another advantage of DBS is that it is relatively easy to include fabrication constraints into the optimisation process. As each pixel represents a physical feature, it ensure that the design can be made according to the minimum feature. 

However, DBS can be computationally expensive, especially for complex designs because it needs to simulate the performance of many different designs in order to find the optimal one. Moreover, DBS can be sensitive to the initial condition. If the initial design is not well chosen, the algorithm may converge to a local optimum. This can be mitigated by running the algorithm multiple times with different initial designs and then selecting the best-performing design. But it is not always possible to guarantee that the global optimum will be found. Numerous methods have been proposed by researchers to improve the computational efficiency of the DBS method \cite{shen2014integrated}.

\begin{figure}[h]
     \centering
     \includegraphics[width=0.6\textwidth]{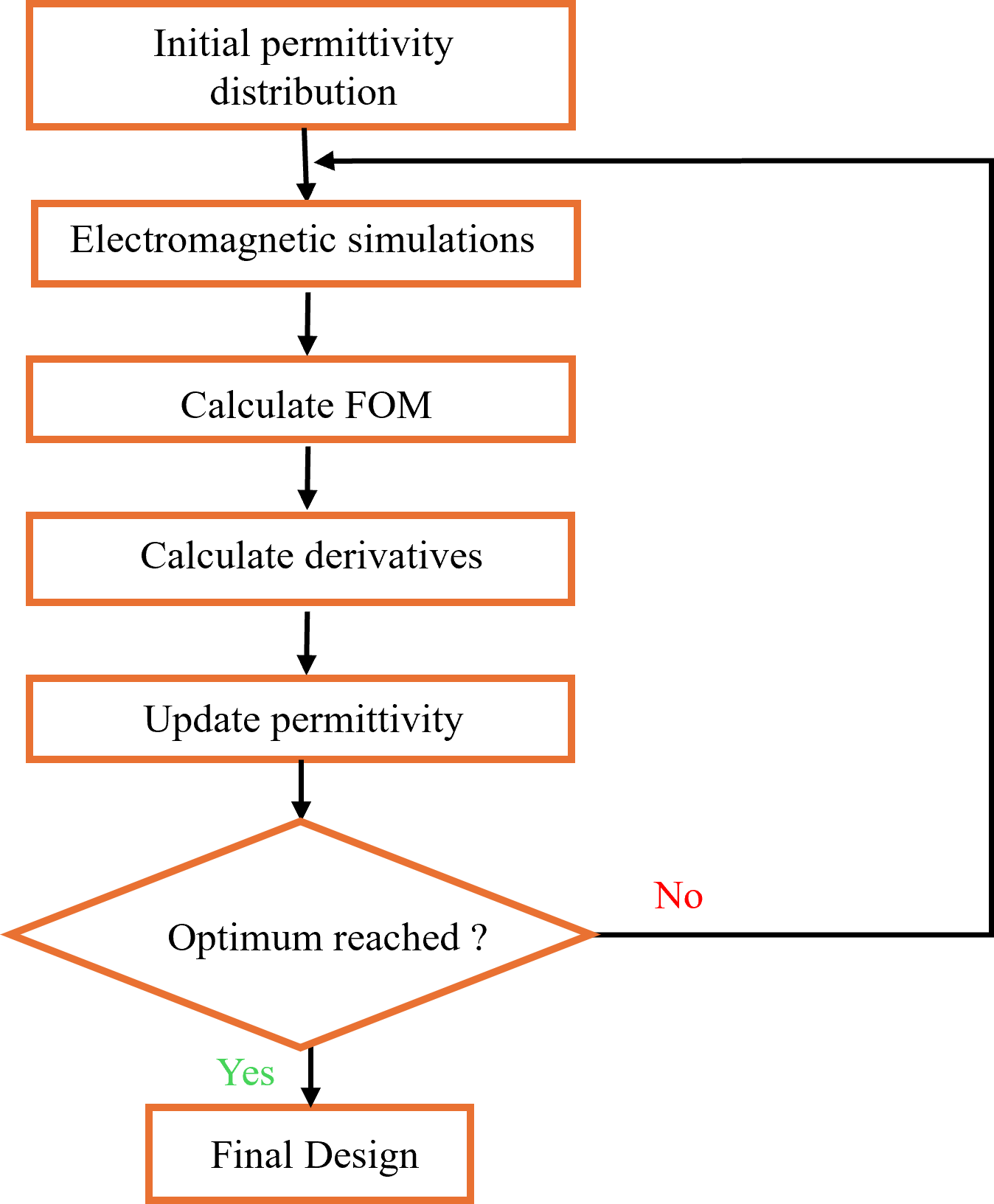}
     \caption{The optimisation process of TO.}
     \label{fig:to_flow}
 \end{figure}
\subsection{Gradient based Topology Optimisation} 

In global search algorithms, the cost to optimize the design increases exponentially with an increase in parameters. Global search algorithms segment the design area into small pixels of dimensions around 10 nm for irregular structures. This leads to a larger problem space due to the higher degree of freedom, which requires huge computation power. Gradient based topology removes this bottleneck and provides an efficient solution for the irregular structures with large parameter spaces. However, each pixel in the design area requires a simulation to calculate its gradient. As the number of design parameters increases, the total calculation time also increases. To overcome these limitations, the adjoint method has been introduced and implemented. Various photonic irregular structures, such as arrayed waveguide grating and ring resonator arrays, are designed and optimized using the adjoint method. The adjoint method enables efficient design of irregular structures with high degrees of freedom (DOF) through the use of gradient-based topology optimisation \cite{jensen2011topology}. This technique has been widely used for shape optimisation in mechanical engineering. The adjoint method computes shape derivatives at every point in space by performing two electromagnetic simulations per iteration.  The adjoint method calculates shape derivatives rapidly and allows for the optimisation of structures by wrapping an inverse algorithm around Maxwell solvers.
The topology optimisation has been used for designing a variety of high-performance photonic devices, including wavelength demultiplexers \cite{su2018inverse}, polarization beam splitters \cite{frandsen2016inverse}, mode converters \cite{frandsen2014topology}, polarization rotator \cite{lebbe2019high, de2022broadband}, all-optical logic \cite{necseli2022inverse}. These examples demonstrate the power and flexibility of topology optimisation in designing a wide array of photonic devices compared to traditional design approaches \cite{shang2023inverse, liang2022topological, piggott2020inverse}.
Pigget et al. developed a silicon wavelength demultiplexer capable of dividing 1300 nm and 1550 nm light from an input waveguide into two output waveguides. They fabricated devices exhibiting low insertion loss wide bandwidths of 100 nm with a very compact footprint of 2.8 $\mu m$ $\times$ 2.8 $\mu m$ \cite{piggott2015inverse}. 

A Y-splitter was designed using topology optimisation by et al \cite{lalau2013adjoint}. The Y-splitter was made on a silicon on insulator (SOI) platform with a 220nm-thick silicon waveguide and silicon dioxide cladding. A minimum radius of curvature of 200 nm was imposed on the design. The design region was the central 2 $\mu$m $\times$ 2 $\mu$m domain. The figure-of-merit was defined as transmission into the fundamental mode of the bent output waveguides, which can be obtained from Poynting vectors. The adjoint method was used to compute the shape derivatives at all points in space for the optimisation, with only two electromagnetic simulations per iteration. The adjoint method is very computationally inexpensive compared to stochastic optimisation methods, and allows the optimisation to be carried out with a very large number of degrees of freedom. The final optimized device achieved an insertion loss of -0.08 dB after 51 iterations.
\begin{figure}[h]
     \centering
     \includegraphics[width=0.9\textwidth]{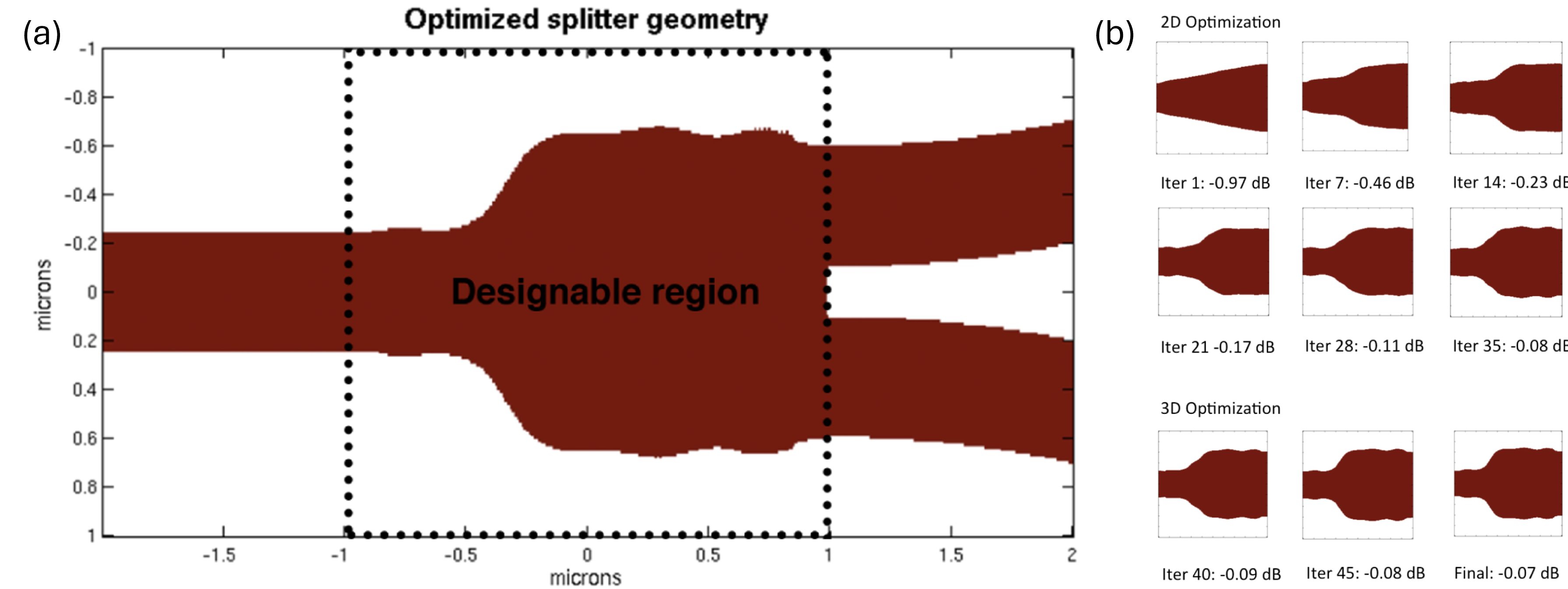}
     \caption{An ultra-compact Y-splitter designed by TO method \cite{lalau2013adjoint}.}
     \label{fig:to_axample}
 \end{figure}

One challenge in topology optimisation is ensuring the manufacturability of the resulting designs. The optimisation process might lead to intricate structures with features smaller than what fabrication techniques can handle. To address this, the filtering technique is used to impose a minimum length scale on the features of the optimized structure, preventing the formation of small features. However, filtering can negatively impact device performance as it limits the design space. There is a trade-off between manufacturability and device performance.

\subsection{Deep Neural Network}  

 Deep neural networks are inspired by the human brain and can recognize complex patterns in text, sound, pictures, and other data types to give accurate insights and predictions. A deep neural network has been used in various computer applications, such as describing images or translating languages. A deep neural network comprises several essential components. An input layer is where data enters the neural network through multiple nodes. Then hidden layers process and transmit data to deeper layers within the network, adapting their responses based on new information. These layers, which can number in the hundreds for deep learning networks, analyze problems from various perspectives. The output layer comprises nodes responsible for producing results.
 
\begin{figure}[h]
     \centering
     \includegraphics[width=0.8\textwidth]{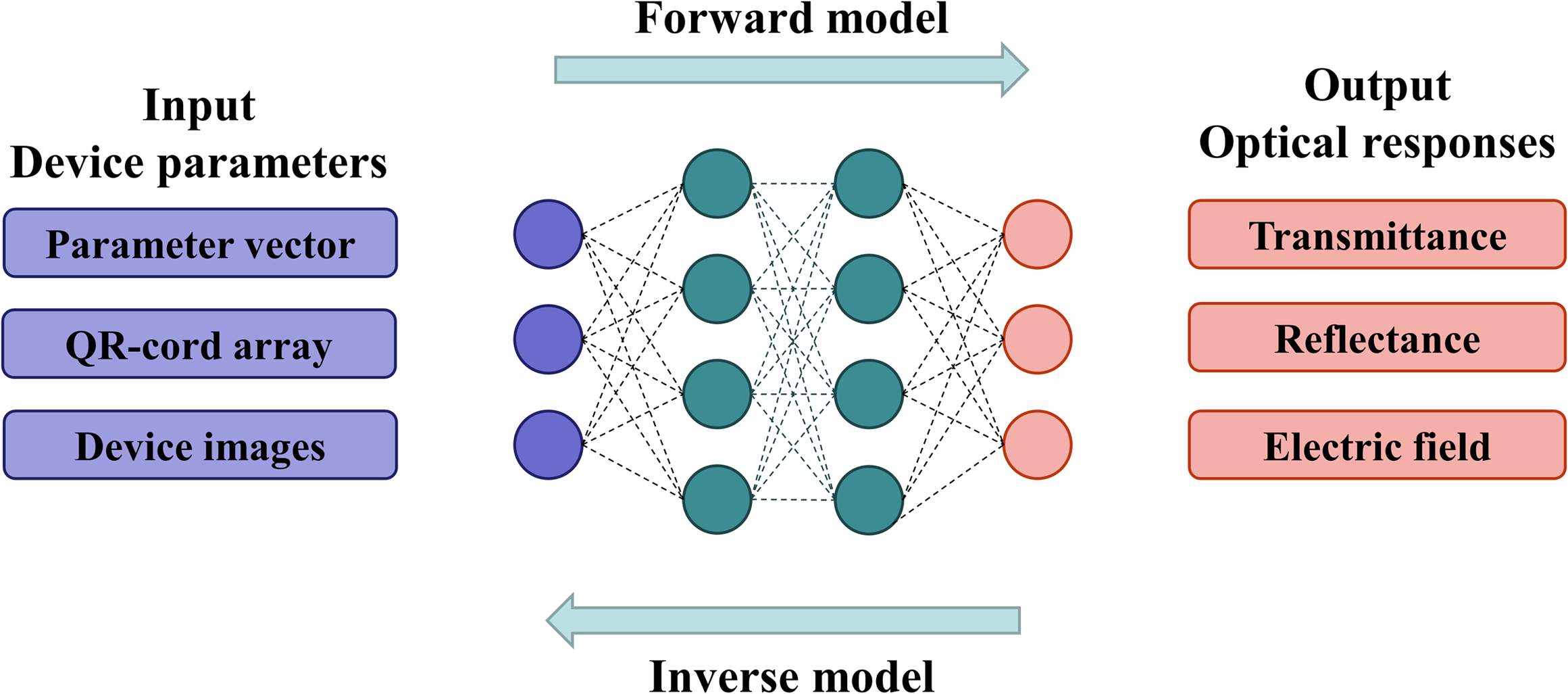}
     \caption{Forward and Inverse model flow for the DNN based inverse optimisation techniques \cite{shen2024recent}.}
     \label{fig:DNN_flow}
 \end{figure}

Recently, DNNs have emerged as a powerful technique for the inverse design and optimisation of photonic devices. DNN offers several advantages over traditional iterative optimisation techniques. It is very useful when designing a series of devices with similar functions but different target outputs. DNNs act as surrogate models for computationally expensive simulations like FDTD or BPM.  Once the DNN model is trained, the DNN can rapidly predict the optical output response for a new set of device geometrical parameters.

With DNNs, photonic device optimisation can be carried out in forward or backward directions. In the forward method, the model is first trained using simulation data to calculate network parameters that minimize loss with the gradient descent algorithm. Subsequently, the forward model can predict the device's response quickly and accurately. These predictions do not necessitate numerical simulations for each prediction; simulations are only required during the model's training phase.

Inverse design based on DNN achieves optimal results using neural networks without relying on traditional inverse design methodologies. The DNN tackles the inverse design problem by learning how to map desired optical responses to corresponding device parameters. However, this approach faces a challenge due to the non-unique nature of inverse problems. As different photonic structures can lead to similar optical responses. Several strategies are employed to eliminate this issue, such as tandem networks \cite{liu2018training}, dimensionality reduction \cite{zandehshahvar2021inverse}, and deep convolutional mixture density networks \cite{unni2020deep}.
\begin{figure}[h]
     \centering
     \includegraphics[width=0.7\textwidth]{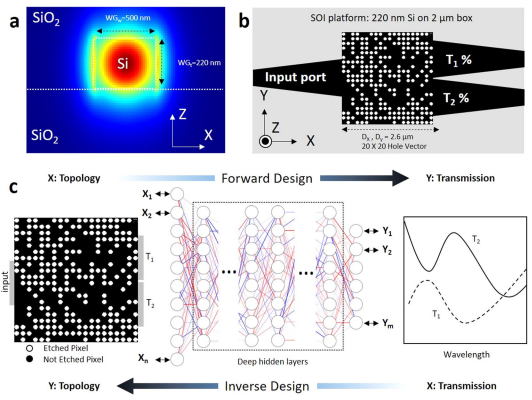}
     \caption{ (a) and (b) An integrated photonic power splitter designed utilizing DNN based prediction. (c) DNN for forward and inverse modeling of nanophotonic devices. The DNN models can take device topology design as input and spectral response of the
metadevice as label or vice versa \cite{tahersima2019deep}.}
     \label{fig:DNN_flow}
 \end{figure}
The primary advantage of DNNs is their ability to enhance inverse design efficiency by eliminating the time-consuming and computationally expensive numerical simulation calculations at each step. Moreover, DNNs can predict novel and highly efficient designs that might not be possible using conventional approaches.

However, DNNs require large and comprehensive datasets for training which are calculated using numerical simulations. Also, the performance of the DNN model can be limited by the quality of the training data. If the training data does not adequately represent the design space, the DNN model might not produce results with high accuracy .

Although DNNs have shown great potential and success for the photonic device design, The DNNs are still largely "black boxes" because their predictions are not easily understood. Despite these challenges, DNNs are powerful means to accelerate the design process and unlock novel device functionalities. Combining DNNs with inverse design methods is an active research area. Further advancements in DNN-based optimisation techniques include the use of physics-assisted neural networks, which reduce the need for precalculated datasets obtained through numerical calculations and enhance data efficiency \cite{chen2020physics}. Another DNN optimisation method, reinforcement learning, has been employed in the design of novel nanophotonic devices \cite{seo2021structural}. Although this method requires substantial computing resources for network training, it does not rely on precomputed datasets.

\subsubsection{Photontorch - Blending Simulation and Optimization}

Photontorch is an innovative simulation and optimization framework built on PyTorch that models photonic circuits as sparse, complex-valued neural networks. By integrating the scatter matrix (S-matrix) formalism with ML techniques, Photontorch enables efficient, parallelized simulation and optimization of PICs~\cite{laporte2019photontorch,laporte2019highly}. Importantly, as open-source, it allows customization and extension to meet specific design needs, making it an adaptable tool for a wide range of applications.

A defining feature of Photontorch is its ability to treat photonic circuits as recurrent neural networks (RNNs). This approach leverages GPU acceleration to efficiently simulate large-scale circuits, such as Coupled Resonator Optical Waveguides (CROWs). Through backpropagation, Photontorch optimizes design parameters directly, avoiding the computational expense of traditional iterative techniques. It has demonstrated its utility by rapidly configuring devices like bandpass filters and programmable photonic networks~\cite{laporte2019photontorch}.

In comparison to commercial tools like Ansys’ Lumerical INTERCONNECT, Photontorch offers significant advantages. For example, it facilitates the frequency-domain characterization of microring resonator (MRR) weight banks, similar to INTERCONNECT ~\cite{lumerical}, but benefits from the accelerated matrix multiplications provided by PyTorch. This results in much faster simulation times. Furthermore, Python’s extensive library ecosystem allows for seamless integration with other system-level simulation tools, enhancing its flexibility~\cite{nichols2023}.

Photontorch has also shown promise in practical applications, such as simulating photonic reservoirs for equalization in optical communication systems. These reservoirs effectively address nonlinear distortions and dispersion effects in high-speed optical links, showcasing the framework’s ability to tackle real-world photonic challenges~\cite{zelaci2024reservoir}. Recent advances in automatic differentiation further underscore the potential for accelerating photonic inverse design workflows, complementing Photontorch’s strengths in gradient-based optimization ~\cite{autodiff2023}.

Last, but not least, its GPU-enabled parallelism makes it well-suited for deployment on high-performance computing (HPC) systems. This scalability can shift photonic circuit simulation from laptop-level computations to HPC environments, greatly accelerating the design process and boosting industrial adoption~\cite{laporte2019highly}, specially when considering large-scale photonic devices~\cite{kang2024large}.

\label{sec3}

\section{Quantum-boosted Design}

Quantum computing, with its unparalleled ability to process information in exponentially large state spaces, introduces novel approaches to tackling optimization problems that are otherwise intractable for classical systems. 

This section explores quantum and quantum-inspired methods as promising alternatives to traditional and machine learning-based approaches, highlighting their unique computational paradigms and their potential for disruptive innovation. 

We present the case for the synergistic relationship between quantum methods and photonic technologies, despite their developmental stage, where each field leverages the other’s advancements to address shared complexities, unlocking scalable and efficient solutions for future innovations.

\subsection{Quantum Algorithms}

The ability of quantum information processing methods to exploit superposition, entanglement, and interference enables fundamentally new approaches to solving optimization problems, many of which are central to PIC design. Quantum algorithms stand out for their capacity to achieve speedups across a range of tasks, and their categorization provides insight into the scale of potential advantages.

\begin{figure}[h!]
    \centering
    \includegraphics[width=0.45\linewidth]{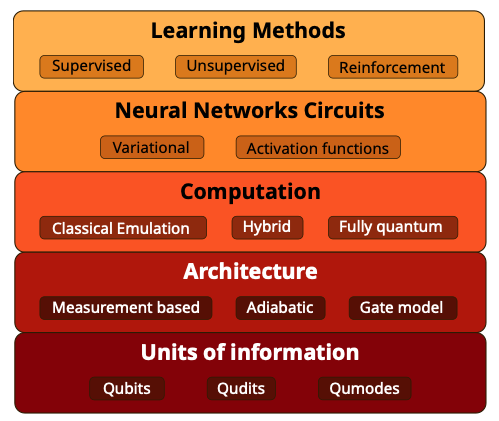}
    \caption{Categorization of topics in quantum machine learning and algorithms, from Prati et al. \cite{qml_speedup_source}}
    \label{fig:qml_categorization}
\end{figure}

To better contextualize quantum optimization techniques, Figure~\ref{fig:qml_categorization} presents a hierarchy of key topics. These include learning methods, neural network circuits, and the types of quantum systems used for computation. The figure highlights the diversity and adaptability of quantum algorithms across various computational architectures.

The breadth of quantum algorithms and their potential to deliver speedups can be found in tasks beyond optimization. Table~\ref{table:quantum_algorithms}, adapted from Biamonte et al. \cite{biamonte2017}, summarizes some of the most notable quantum algorithms and their speedup characteristics, highlighting the interplay between algorithmic advancements and hardware constraints.

\begin{table}[h!]
    \centering
    \begin{tabular}{|l|l|}
    \hline
    \textbf{Methods} & \textbf{Speed-up} \\ \hline
    Bayesian inference & \(O(\sqrt{N})\) \\ \hline
    Online perceptron & \(O(\sqrt{N})\) \\ \hline
    Least-squares fitting & \(O(\log(N))\) \\ \hline
    Classical Boltzmann machine & \(O(\sqrt{N})\) \\ \hline
    Quantum Boltzmann machine & \(O(\log(N))\) \\ \hline
    Quantum PCA & \(O(\log(N))\) \\ \hline
    Quantum support vector machine & \(O(\log(N))\) \\ \hline
    Quantum reinforcement learning & \(O(\sqrt{N})\) \\ \hline
    \end{tabular}
    \caption{Speed-up of quantum machine learning subroutines. Adapted from Biamonte et al. \cite{biamonte2017}}
    \label{table:quantum_algorithms}
\end{table}

While many of these algorithms remain in exploratory phases, their applications to photonic optimization are particularly promising. For example, algorithms such as quantum PCA and quantum reinforcement learning highlight the potential for quantum-assisted techniques in high-dimensional parameter spaces. Nevertheless, realizing these advantages requires overcoming the limitations of current quantum hardware, which often lacks the scale and fault tolerance necessary for full implementation. Despite these challenges, the rapid pace of innovation in both algorithms and hardware signals a promising trajectory for quantum-enhanced photonic design.

In the following sections, we focus on two algorithms that have seen experimental validation and are particularly suited for photonic applications: the Variational Quantum Eigensolver (VQE) and the Quantum Approximate Optimization Algorithm (QAOA). By encoding the optimization problem into a quantum system, these algorithms iteratively adjust quantum gate parameters to find the optimal solution with fewer computation steps. This process requires components that include a parameterized trial solution and problem-specific information to enhance the structure of the ansatz circuit. Both have extensively demonstrated their utility in addressing complex optimization problems in photonic platforms \cite{zhong2020}.

\subsubsection{Variational Quantum Eigensolver}

The VQE is a hybrid quantum-classical algorithm. It seeks to minimize the expectation value of a Hamiltonian \( H \), which represents the physical properties or cost functions of the system, such as loss, coupling efficiency, or nonlinear interaction strengths. The minimization is expressed as:
\[
E(\theta) = \langle \psi(\theta) | H | \psi(\theta) \rangle,
\]
where \( |\psi(\theta)\rangle \) is a parameterized quantum state (ansatz) defined by tunable parameters \( \theta \). 

The iterative process of VQE begins by preparing \( |\psi(\theta)\rangle \) on a quantum processor. Measurements of \( E(\theta) \) are then performed, and the results are used in a classical optimization loop to update \( \theta \). This cycle continues until convergence to the minimum eigenvalue of \( H \), which corresponds to the system's optimal configuration. This hybrid approach enables VQE to efficiently navigate high-dimensional and non-convex landscapes, leveraging quantum superposition and entanglement to explore multiple configurations simultaneously. VQE has demonstrated robustness to experimental noise, making it a practical tool for noisy intermediate-scale quantum (NISQ) devices.

\begin{figure}[h!]
    \centering
    \includegraphics[width=0.9\linewidth]{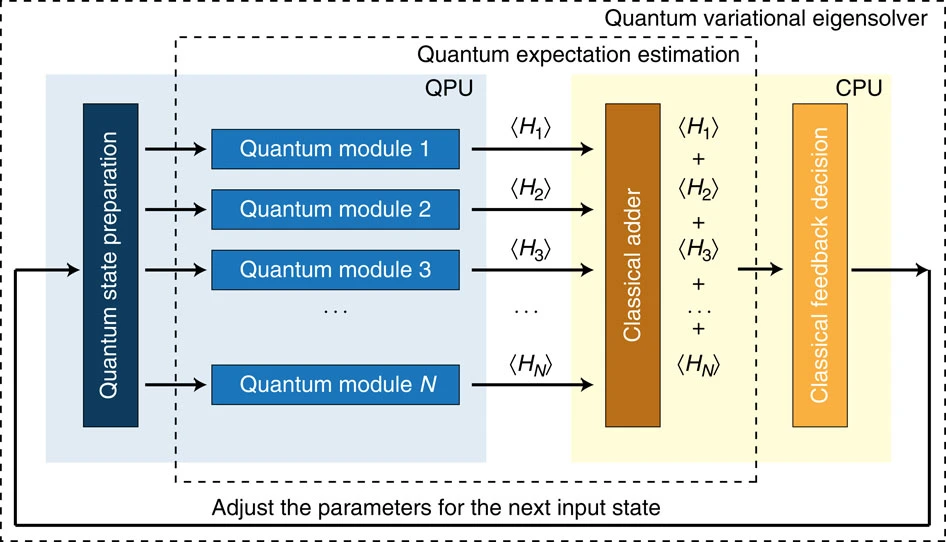}
    \caption{Schematic representation of the Variational Quantum Eigensolver (VQE) workflow. The QPU prepares quantum states and estimates expectation values, which are then processed by the CPU to optimize the parameters for the next iteration \cite{peruzzo2014}.}
    \label{fig:vqe_diagram}
\end{figure}

Figure~\ref{fig:vqe_diagram} provides a schematic overview of the VQE workflow, illustrating the hybrid quantum-classical process. The quantum processor (QPU) prepares the quantum state, estimates the expectation values of the Hamiltonian, and sends the results to the classical processor (CPU) for optimization and parameter updates. This iterative feedback loop continues until the system reaches convergence.

VQE has been used successfully in theoretical demonstrations, particularly in quantum chemistry, where it computes molecular ground-state energies with significantly fewer resources than traditional diagonalization methods. For example, simulations have achieved chemical accuracy for molecules such as \( \text{H}_2 \) and HeH\(^+\), highlighting the algorithm's feasibility under noisy conditions \cite{bentellis2023, gard2019}. 

In photonic device design, \( H \) can represent specific objectives. For example, in waveguide networks, \( H \) encapsulates propagation losses originating from internal material absorption, sidewall roughness, and bending-induced radiation, which can be derived as imaginary (loss) terms from Maxwell’s equations in the paraxial approximation. The parameterized quantum state \( |\psi(\theta)\rangle \) corresponds to the complex amplitudes and phases of electromagnetic fields in these waveguides. By iteratively adjusting \( \theta \), VQE explores the design space to minimize propagation losses while preserving optical signal integrity (e.g., reflectance, transmission).

Unlike purely classical inverse-design approaches, VQE employs a quantum-inspired parameter search that can access regions of the solution space that are difficult to reach with standard gradient-based or heuristic algorithms.

Let us see how this framework can be applied in practice. For nonlinear photonic devices, such as frequency converters, \( H \) can encode the efficiency of nonlinear interactions. For instance, in second-harmonic generation (SHG), the optimization process involves maximizing the overlap integral between interacting modes and achieving precise phase matching conditions. By tailoring ferroelectric domain structures in lithium niobate (LiNb) waveguides or fine-tuning resonator coupling parameters, studies have demonstrated significant enhancements in SHG efficiency \cite{shen2024, arxiv2024}.

The hybrid structure of VQE—combining the computational power of quantum algorithms with the adaptability of classical optimization—ensures its scalability alongside advances in quantum hardware. This positions VQE as a transformative tool for photonic device optimization. Its ability to incorporate physical constraints, explore high-dimensional parameter spaces, and handle noisy environments paves the way for breakthroughs in optical interconnects, quantum communication networks, and photonic neural networks \cite{li2019, tilly2021}.

\subsubsection{Quantum Approximate Optimization Algorithm}

\begin{figure}[h!]
    \centering
    \includegraphics[width=\linewidth]{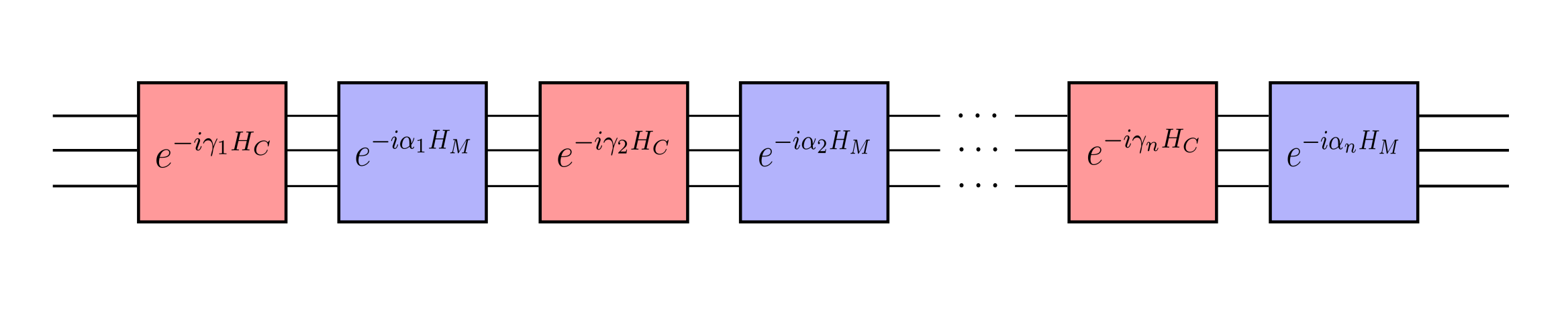}
    \caption{Circuit representation of QAOA. Each layer alternates between applying the cost Hamiltonian (\(e^{-i\gamma H_C}\)) and the mixer Hamiltonian (\(e^{-i\beta H_M}\)), optimizing the parameters \((\gamma, \beta)\) iteratively. Adapted from the tutorial by PennyLane \cite{pennylane_qaoa_tutorial}. For more details, refer to the online resource.}
    \label{fig:qaoa_diagram}
\end{figure}

The QAOA is a hybrid quantum-classical framework for tackling combinatorial optimization problems. As depicted in Figure~\ref{fig:qaoa_diagram}, it alternates between applying the problem Hamiltonian \(H_C\) (encoded as \(U_C\)) and a mixing Hamiltonian \(H_M\) (encoded as \(U_M\)). In each iteration, the variational parameters \((\gamma, \beta)\) are updated via a classical feedback loop to minimize the cost function encoded in \(H_C\). The quantum state evolves as:
\[
|\psi_p\rangle = \prod_{k=1}^p e^{-i \beta_k H_M} e^{-i \gamma_k H_C} |\psi_0\rangle,
\]
where \( |\psi_0\rangle \) is the initial state, and \( \gamma_k, \beta_k \) are variational parameters optimized using a classical optimizer.

Unlike traditional design or machine learning (ML) approaches, QAOA systematically encodes problem constraints into a cost Hamiltonian, which can yield better convergence and more robust solutions. For instance, in Routing and Spectrum Assignment (RSA) problems for Elastic Optical Networks (EONs), QAOA minimizes delay while adhering to spectrum continuity constraints \cite{bouchmal2024_photonics}. Similarly, it has demonstrated potential in 6G optical backhaul and metro networks, where it efficiently handles latency and throughput requirements \cite{bouchmal2024_6g}.

The ability of QAOA to encode and explore large design spaces extends to photonic device design and optimization. In waveguide networks, for example, the topology can be represented as a graph, where nodes correspond to waveguides and edges encode properties like propagation loss or crosstalk. The cost Hamiltonian \( H_C \) models total losses, while the mixer Hamiltonian \( H_M \) explores alternative configurations, identifying routing schemes that minimize losses and interference \cite{zhou2018, shaydulin2019}. This graph-based representation enables QAOA to optimize complex waveguide arrangements effectively while adhering to design constraints such as minimum waveguide separation, restricted areas, and bending radii, as specified by manufacturing foundries.

For programmable photonic circuits, QAOA can optimize phase shifter placements and tunings to achieve precise interference patterns, which are critical in applications like quantum communication and photonic computing. Similarly, in resonator-based devices such as ring resonators, QAOA can tune coupling coefficients to maximize Q-factors or minimize back-reflection, enhancing overall device performance \cite{enomoto2022, farhi2014}. These examples illustrate the adaptability of QAOA to a variety of photonic device optimization challenges.

By integrating foundry-defined rules, such as bending radii and restricted areas, into the cost Hamiltonian, QAOA ensures compliance with practical manufacturing constraints. Further evidence of its potential can be found in \cite{bouchmal2024_photonics}, where QAOA outperformed traditional methods like Integer Linear Programming (ILP) and Deep Reinforcement Learning (DRL) for solving the RSA problem.

While current quantum hardware imposes limits on solution quality, QAOA’s scalability and adaptability underscore its potential for addressing optimization tasks that demand rapid and robust decision-making. By leveraging quantum parallelism and hybrid optimization, QAOA provides a systematic and scalable framework for balancing performance metrics (e.g., insertion loss, coupling efficiency) with fabrication tolerances. As quantum hardware advances, QAOA’s integration into photonic design workflows promises to drive innovation in optical interconnects, quantum networks, and beyond \cite{akshay2021, hadfield2018}.

\subsection{Quantum-Inspired Methods}
The design and optimization of PIC increasingly demand accurate, large-scale simulations before committing to costly fabrication steps. Traditional computational techniques face significant challenges in modeling the complex quantum-level interactions that dictate device performance. While quantum computing offers a promising avenue to tackle such complexity, the direct application of quantum hardware to PIC optimization remains elusive. Current quantum devices suffer from high error rates, and translating practical PIC design problems into suitable quantum algorithms presents its own difficulties.

To address these challenges, hybrid quantum-classical approaches have emerged. By integrating quantum-inspired techniques—particularly tensor network methods—with classical optimization routines, we can leverage the rich representational capacity of quantum-inspired frameworks to explore a vast design space. Classical optimizers then refine these candidate solutions to meet real-world fabrication and performance requirements. Beyond mere optimization, this synergy offers a “quantum boost” in simulation fidelity, enabling more accurate and reliable PIC modeling. As a result, more efficient design iterations can accelerate the development cycle of the next-generation of PIC \cite{sparrow2019, somhorst2023}.

\subsubsection{Introduction to Tensor Networks}

\begin{figure}[h]
     \centering
     \includegraphics[width=0.7\textwidth]{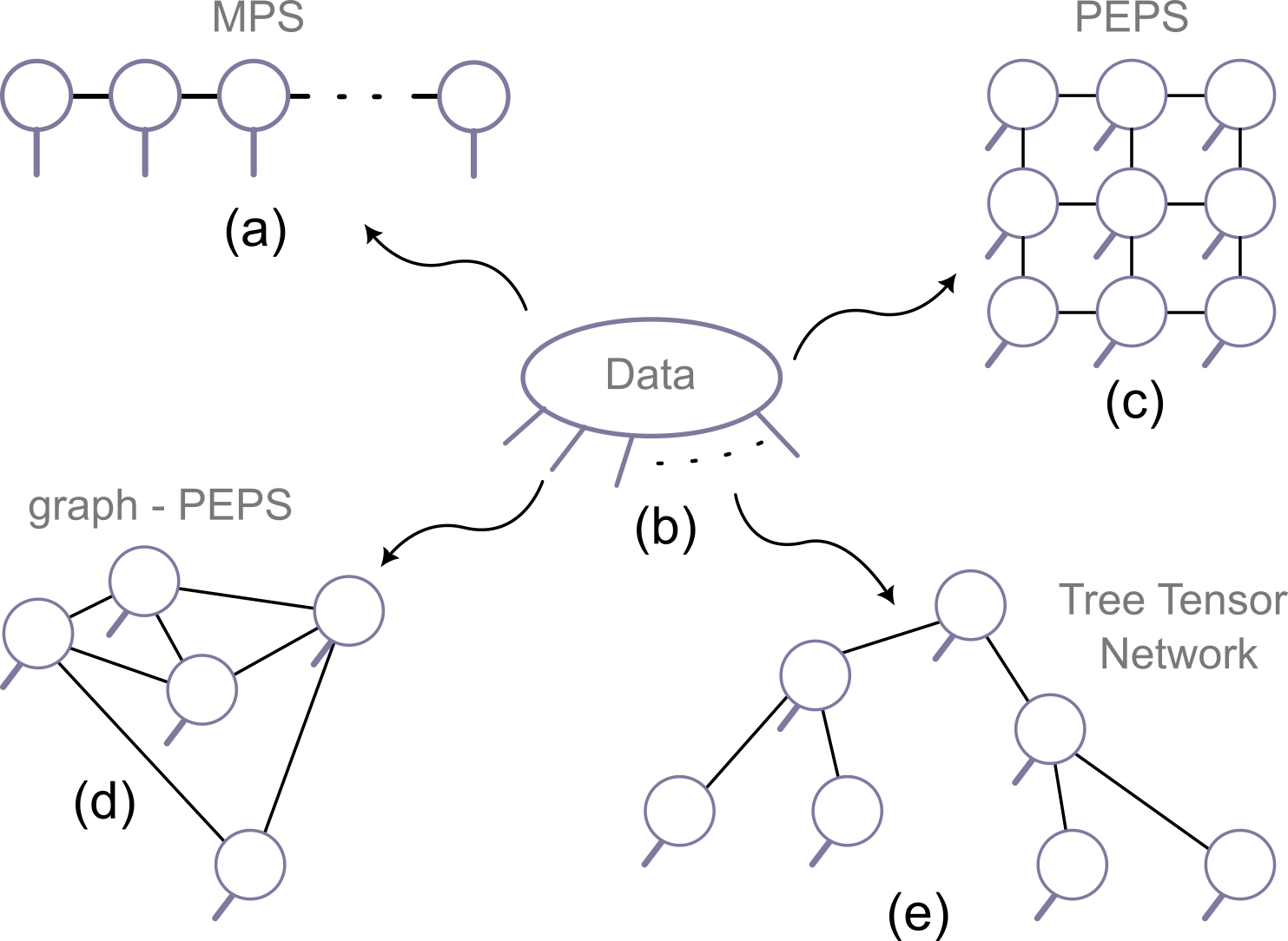}
     \caption{Different tensor network ansatz for efficiently representing higher-dimensional data (b). Examples include (a) Matrix Product State (MPS) or tensor train, (c) a two-dimensional Projected Entangled Pair State (PEPS), and (d) more general structures such as graph-based PEPS and (e) Tree Tensor Networks.}
     \label{fig:TNs}
 \end{figure}

The exponential growth of the memory footprint of a data structure as the number of variables increases poses a formidable challenge across many scientific domains, from quantum many-body physics to machine learning. For instance, representing the wave-function of an $n$-qubit system requires exponentially large ($2^n$) parameters, a quantity that becomes prohibitively large even for moderate $n$. Yet, not all parameters are equally crucial; in most practical scenarios, the system’s core properties can be captured with polynomial number of parameters due to limited entanglement or other structural constraints. By exploiting this inherent sparsity and reorganizing the representation around the parameters that matter most, we can drastically reduce memory requirements. This allows us to preserve the accuracy needed for practical applications, while avoiding the computational intractability that arises from brute-force methods.

Tensor networks \cite{PEPSReviewOrus,PEPSReviewOrus2,PEPSReviewCirac} provide an efficient framework for representing high-dimensional data. Instead of encoding the entire dataset or quantum state as a single, exponentially large tensor, tensor networks hold the relevant information in a network of smaller, interconnected tensors. These smaller tensors capture local correlations within the data, while their connections represent entanglement and dependencies between different parts of the system. This decomposition significantly reduces the computational complexity required for representation and future manipulation.

Beyond providing compact representations, tensor networks come equipped with a suite of powerful techniques for manipulating and analyzing complex systems. Methods like Time-Evolving Block Decimation (TEBD) \cite{TEBD1,TEBD2} enable efficient simulation of time-dependent dynamics, while the Density Matrix Renormalization Group (DMRG) \cite{DMRG} algorithm excels at finding low-energy states in strongly correlated systems. These foundational approaches introduced key tensor network architectures, such as Matrix Product States (MPS) \cite{MPS1,MPS2}, which efficiently represent one-dimensional quantum systems with limited entanglement. As researchers pushed toward higher-dimensional problems, Projected Entangled Pair States (PEPS) \cite{PEPS} and their generalizations (gPEPS \cite{gPEPS}, Flexible-PEPS \cite{Flexible-PEPS}) were developed to capture complex correlations across two-dimensional and arbitrary geometries respectively.

Tensor networks have since evolved into versatile tools that span multiple disciplines, extending their influence well beyond their origins in condensed matter physics. In quantum computing, they provide a robust framework for simulating quantum circuits \cite{ibm_simulator,TN_attack}, offering critical insights into algorithmic performance and error resilience in noisy environments. In artificial intelligence (AI) and machine learning \cite{TN_in_ML,TN_in_ML2,TN_in_ML3}, tensor networks aid in compressing large models \cite{CompactifAI}, making high-dimensional data analysis more tractable. Their utility further extends into areas such as data science \cite{TN_in_data_science}, string theory \cite{TN_in_string_theory}, category theory \cite{TN_in_category_theory}, and mathematical physics \cite{TN_in_math_physics}, illustrating their broad applicability. The following section delves deeper into one particularly important application: simulating quantum circuits using tensor networks.

\subsubsection{Quantum-inspired quantum circuit simulator }
\begin{figure}[h]
     \centering
     \includegraphics[width=0.9\textwidth]{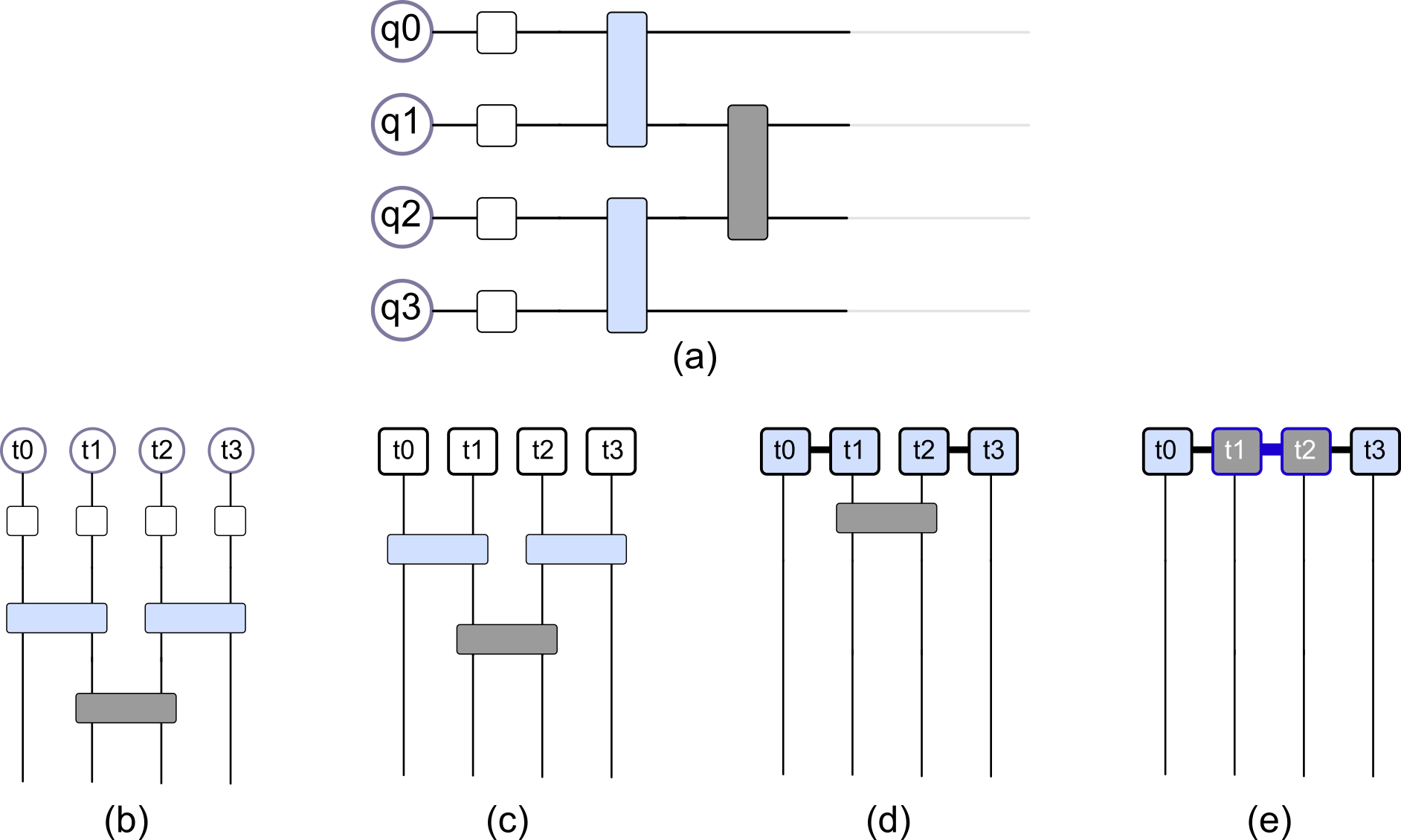}
     \caption{(a) A quantum circuit of four qubits, each subjected to a few single- and two-qubit gates. All qubits are initialized in the state $|0\rangle$, and time progresses from left to right. (b) The corresponding tensor network representation. Here, each initial qubit is replaced by an order$-1$ vertex tensor (a vector), initialized as $(1, 0)^T$. (c) The exact absorption of a single-qubit gate into the vertex tensors. (d) The absorption of a two-qubit gate updates the vertex tensors and creates entanglement between the involved qubits, represented by a bold horizontal edge connecting the two vertices. (e) Additional two-qubit gates further increase the entanglement within the system.}
     \label{fig:QC_TN}
 \end{figure}

\textit{Simulating quantum circuits:} Simulating quantum circuits on classical computer is essential for validating, benchmarking, and exploring quantum algorithms and hardware. Such circuits consist of sequences of quantum gates acting on qubits, generating entangled state. Direct brute-force methods quickly become infeasible, where tensor networks provide a scalable and efficient framework to address this challenge by exploiting the limited entanglement that often characterizes realistic quantum computations. In this approach, quantum states and operations are represented as interconnected tensors that encode local correlations and entanglement patterns. Mapping a quantum circuit onto a tensor network (Fig.\ref{fig:QC_TN}) involves expressing the initial state as a product of simple tensors and then applying each gate as a series of tensor contractions. This construction closely mirrors the logical flow of the original circuit, enabling simulations of state evolution and final measurements. 

A key advantage of tensor network simulations is their capacity to manage entanglement growth. In systems where entanglement remains limited, tensor networks maintain compact representations, and the computational cost depends primarily on the bond dimension—the size of the indices linking adjacent tensors. For one-dimensional circuits or those with limited connectivity, Matrix Product States (MPS) provide an optimal representation. MPS representation is particularly advantageous for photonic circuits with large numbers of modes and photons. By combining the S-matrix formalism with MPS, simulations can achieve high efficiency while retaining accuracy, making this approach ideal for scalable quantum photonic simulations. However, when circuits exhibit more complex, higher-dimensional connectivity, Projected Entangled Pair States (PEPS) offer a more accurate and scalable approach.

A key factor enabling efficient quantum circuit simulation is the ability to accurately handle two primary computational tasks :

\begin{itemize} 
\item \textbf{Gate Application:} Single-qubit gates are directly applied through exact tensor contractions, efficiently updating the local tensors that represent the targeted qubit. For two-qubit gates, a simple-update variant of the Time-Evolving Block Decimation (TEBD) algorithm is employed, allowing for controlled approximations that maintain tractability as the circuit grows in depth. 
\item \textbf{Measurements:} During measurements, only local environment tensors are initially considered, enabling direct calculation of probabilities and expectation values associated with individual qubits. More comprehensive schemes, which involve contracting larger tensor clusters or even the entire tensor networks, become possible at higher computational costs. This spectrum of approaches creates a hierarchy of measurement strategies, allowing one to balance accuracy with available computational resources.
\end{itemize}
These capabilities ensure that tensor network simulators can effectively handle a broad range of operations, from simple one-qubit gates to intricate two-qubit interactions. Beyond conventional unitary gates, tensor networks can also incorporate non-unitary transformations—such as those associated with noise or certain error-correction protocols—expanding their applicability to more realistic and challenging quantum circuit simulations.

Despite these strengths, tensor network simulations encounter difficulties when circuits exhibit extensive entanglement or complex connectivity. Under these conditions, bond dimensions—and thus computational overhead—can become prohibitively large, even with aggressive bond truncation. As a result, carefully balancing accuracy and feasibility becomes crucial. Tensor networks have already demonstrated their value across a wide range of quantum circuit simulations \cite{TN_qc_simulator,TN_qc_simulator2}. In the following section, we present a concrete application: using tensor networks to simulate IBM’s advanced noise-less quantum processors. This case study illustrates the scalability and effectiveness of tensor network methods in tackling large-scale quantum circuits, showcasing their potential to guide the future of quantum computing research and development.

\textit{Simulating IBM's kicked Ising experiment:}
IBM’s ongoing advancements in superconducting quantum processors, including the Eagle (127 qubits), Osprey (433 qubits), and Condor (1121 qubits) architectures, provide increasingly large testbeds for quantum computational tasks. However, the growing complexity of these devices also presents significant challenges for benchmarking and validation. In this context, tensor network simulations have proven instrumental, offering a powerful tool to simulate quantum circuits at scales where direct classical methods falter and hardware noise limits the fidelity of experimental results.

By adopting a PEPS representation that closely matches the two-dimensional arrangement of IBM’s processors, our simulations of the Kicked Ising experiment—a real-time dynamics study replicating complex circuit evolution originally investigated using IBM’s QPU \cite{IBM_nature}—achieved a high-fidelity reconstruction of the target quantum states and observed measurement outcomes. This approach was effective even when simulating circuits at sizes that push the boundaries of contemporary hardware capabilities. Such results validate the underlying hardware design and associated quantum protocols, while also serving as a litmus test for emerging techniques in error mitigation and circuit optimization. This PEPS based-quantum circuit simulator not only outperforms IBM's QPU in terms of both accuracy and speed but also demonstrates that PEPS-based approaches significantly surpass IBM's MPS-based and other PEPS-based simulation results \cite{IBM_tn_tindall}.

Extending beyond Eagle, scaling tensor network simulations to Osprey and Condor has demonstrated the feasibility of handling circuits at the thousand-qubit scale. By simulating systems with eight times the more qubits and twice longer time evolution, we have showcased the power of our PEPS-based quantum circuit simulator. These advances not only guide the refinement of device architectures but also establish benchmarks for future quantum algorithms and large-scale quantum processors.

\subsubsection{Tensor network simulator in PIC optimization}
As quantum hardware advances toward practical implementations, accurately modeling noise becomes increasingly critical. Real devices, whether photonic or superconducting, operate under non-ideal conditions such as photon loss, mode mismatch, dephasing, decoherence, and gate imperfections, all of which degrade the fidelity of quantum operations. To realistically simulate these systems, incorporating noise processes into tensor network models is essential for reflecting real-world performance, guiding hardware improvements, and validating error mitigation strategies. Common noise models in photonic quantum circuits include photon loss, dephasing, mode-mismatch errors, thermal noise or background light, phase-insensitive amplifier noise, depolarizing channels, and imperfections in optical components like lossy beam splitters and interferometers. For example, photon loss in a single-photon mode, represented by a single-qubit subsystem, can be modeled using the Kraus operator formalism. Given a single-qubit state $\rho$, the photon loss channel $\mathcal{E}$ is represented by Kraus operators ${K_0, K_1}$ acting on $\rho$:
\begin{eqnarray}
K_0 &=& \begin{pmatrix}
1 & 0 \\
0 & \sqrt{1-p}
\end{pmatrix}, ~
K_1 = \begin{pmatrix}
0 & \sqrt{p} \\
0 & 0
\end{pmatrix}.
\end{eqnarray}

Here, $ p $ is the probability of losing a photon. The noisy state is then given by $\rho' = \sum_j K_j \rho K_j^\dagger.$ In a tensor network simulation, these Kraus operators are inserted at the appropriate step, for example, immediately after applying a unitary gate to the qubit. Two-qubit noise can be introduced as noisy two-qubit gates. By the introduction of such custom noises at multiple points in the circuit, one can flexibly study how various noise processes affect the evolving quantum state throughout the computation.

By simulating photonic circuits under realistic noise conditions, we gain valuable insights into the interplay between circuit design and hardware imperfections. This capability empowers researchers and engineers to identify critical bottlenecks and optimize device parameters—such as mode structure, squeezing parameters, interferometer layouts, and coupling efficiencies—to achieve higher-fidelity operations. Consequently, improved circuit configurations can be tested virtually before committing to fabrication and experimentation, thereby reducing the development cycle and guiding the evolution of photonic quantum technologies.

While our primary focus is on photonic systems, the same principles are broadly applicable across different quantum hardware platforms. For instance, superconducting qubit processors encounter challenges such as gate inaccuracies, leakage, and crosstalk errors, which can be modeled using similar noise frameworks. By incorporating these noise effects into tensor network simulations, we can benchmark error rates, evaluate the impact of enhanced gate calibrations, and assess the effectiveness of noise-aware protocols. Ultimately, this comprehensive, noise-inclusive approach facilitates the rational design of next-generation quantum hardware, irrespective of the underlying physical platform.

\label{sec4}

\section{Applications}

The advancement of photonic integrated circuits heralds a new era in technology, with far-reaching implications across various domains. As we explore the potential of quantum photonic circuits, their ability to overcome traditional computing limitations, and their application in defense, quantum sensing, and space, it becomes evident that PICs are not just theoretical marvels but instrumental in shaping the future of technology.

\subsection{Artificial Intelligence}




The rapid development and scaling of PICs have enabled both industry and academia to explore their potential for providing substantial computational speed-ups while reducing resource consumption.

On the industrial front, companies like Lightmatter and Lightelligence are leading advancements in photonic AI hardware. Lightmatter has developed hybrid programmable processors (Passage) and machine learning accelerators (Envise), achieving faster computations with significantly reduced power consumption, supported by their proprietary deep learning frameworks (Idiom) \cite{lightmatter_website}. Similarly, Lightelligence produces optical equivalents of traditional electronic components, such as multiply-accumulate units and network-on-chip architectures, achieving energy efficiencies as low as 1 pJ/bit \cite{lightelligence_website}.

Complementing these industrial innovations, academic research has made critical strides in optimizing PIC performance for practical AI applications. Xu et al. \cite{xu2022} demonstrated high-order tensor processing on integrated photonic circuits, optimizing for noise resilience to ensure reliable operations. Shi et al. \cite{shi2020} numerically simulated indium phosphide (InP) photonic cross-connects, highlighting significant improvements in energy efficiency when tailored for deep neural networks.

PICs inherently address bottlenecks in traditional electronic systems by enabling high-bandwidth, low-latency communication essential for modern AI workloads. This capability is particularly beneficial for large-scale models, such as transformers, where computational complexity requires hardware that can handle massive data throughput efficiently \cite{shastri2021}. Furthermore, wavelength-division multiplexing in photonic systems facilitates parallel matrix multiplications, significantly accelerating core operations in neural networks while minimizing energy use \cite{ning2024}.

The integration of photonic components into compact and energy-efficient systems has also opened opportunities for distributed AI and edge computing. Applications in IoT, wearable technology, and remote sensing benefit from PIC-based hardware, which combines high performance with low power consumption, expanding the scope of AI deployment.

As PIC technology matures, its role in driving sustainable and scalable AI computation will continue to grow, bridging the gap between increasing computational demands and energy efficiency.

\subsection{Quantum Computing}

Photons, as quantum carriers, offer a compelling pathway toward scalable and practical quantum computing. Unlike superconducting qubits, which require cryogenic temperatures and complex infrastructure, photonic qubits operate at room temperature, eliminating the need for dilution refrigerators and extensive cabling. This operational simplicity reduces both the cost and complexity of quantum systems, positioning photonics as a promising platform for achieving quantum advantage \cite{nature_s42254_021}.

Integrated photonic platforms further enhance these benefits by providing a scalable and reconfigurable architecture for quantum algorithms. Companies like Quandela and Xanadu have capitalized on these advantages: Quandela is known for its high-precision single-photon sources and open-source quantum software, Perceval, while Xanadu integrates its quantum hardware with robust cloud platforms, making quantum technologies accessible to a broader audience. Their innovations exemplify the growing ecosystem around photonic quantum computing.

Recent advancements in photon source technology have led to high-quality, indistinguishable single-photon generation, which is crucial for the implementation of quantum algorithms such as the Variational Quantum Eigensolver (VQE) and the Quantum Approximate Optimization Algorithm (QAOA). The VQE has been experimentally demonstrated on photonic processors to solve quantum chemistry problems, such as calculating the ground-state energies of molecules like He–H$^+$ \cite{peruzzo_ncomms_2014}. This hybrid quantum-classical approach leverages the strengths of both systems, enabling efficient simulations of molecular Hamiltonians and fostering applications in materials science and drug discovery.

Similarly, the QAOA has been implemented on programmable photonic quantum computers to address continuous optimization problems. Enomoto et al. demonstrated a continuous-variable version of QAOA on a single-mode photonic quantum processor, highlighting the potential of photonics in solving complex optimization tasks \cite{enomoto_fio_2022}.

The inherent low decoherence rates of photons, due to their weak interaction with the environment, make photonic systems robust against noise, which is advantageous for fault-tolerant quantum computing. Moreover, the compatibility of integrated photonics with existing semiconductor fabrication techniques facilitates the development of compact and scalable quantum devices, accelerating the transition from experimental setups to practical applications \cite{nature_photonics_2021}.

As the field advances, innovations in integrated photonics are expected to drive significant progress in quantum simulation, optimization, and secure communication. The combination of room-temperature operation, scalability, and integration with classical photonic technologies underscores the potential of photonic quantum computing to overcome the limitations of current quantum systems and achieve practical quantum advantage.

\subsection{Defense and Quantum Sensing}

In the realm of defense, PICs are revolutionizing secure communications through quantum key distribution (QKD), a technique that leverages the principles of quantum mechanics to ensure unbreakable encryption. Furthermore, quantum sensing, facilitated by PICs, offers unprecedented sensitivity and precision in measurements, critical for navigation, timing, and surveillance applications, enhancing the strategic capabilities of defense systems \cite{pirandola2020}.

\subsection{Space Communication}

PICs inherent robustness, lightweight, and low energy consumption make them ideal for deployment in satellites and space probes. Quantum photonic circuits enable secure quantum communication links between earth and space assets, paving the way for a global quantum network and enhancing data transmission capabilities for deep space missions \cite{chen2021}.



\label{sec5}

\section{Conclusion}

As photonic integrated circuits scale in complexity and functionality, the limitations of classical design methods are becoming increasingly apparent. While classical techniques have been instrumental in the development of PICs, their reliance on iterative parameter sweeps and heuristic optimization struggles to meet the demands of modern, large-scale circuits. The exponential growth in design complexity, especially in high-dimensional parameter spaces, exposes the inefficiencies of these approaches, making them increasingly impractical for rapid prototyping and optimization. This is particularly evident in densely packed waveguide networks and interferometric circuits, where classical methods often lack scalability and universality.

Quantum methods, such as the Variational Quantum Eigensolver (VQE) and the Quantum Approximate Optimization Algorithm (QAOA), offer a powerful extension to address these challenges. By leveraging quantum properties like superposition and entanglement, these hybrid quantum-classical techniques can efficiently explore vast design spaces, overcoming the computational bottlenecks faced by classical approaches. These methods not only provide scalable solutions for optimizing large and intricate PIC systems but also adapt to the increasing demands of real-time adaptability and high precision, ensuring robust and efficient designs.

The need for quantum methods becomes even more pressing with the growing role of PICs in quantum technologies. Applications such as quantum communication, quantum sensing, and quantum computing require PICs that are tightly integrated with advancements in quantum hardware. However, a significant mismatch often exists between the capabilities of classical PIC design frameworks and the requirements of quantum hardware, resulting in suboptimal designs and lengthy redesign processes. Bridging this gap is essential for ensuring seamless translation of theoretical designs into practical implementations.

Integrating quantum methods into PIC design not only addresses the limitations of classical techniques but also minimizes mismatches with quantum hardware development. This alignment ensures that PICs remain at the forefront of enabling both classical and quantum technologies. By embracing this transition, PIC design can achieve the universality, scalability, and efficiency required to support the next generation of photonic systems.\label{sec6}




\backmatter

\newpage

\bibliography{bibliography}

\newpage

\begin{appendices}


\appendix
\section{Mathematical Formulations of Simulation Techniques}

\subsection{Beam Propagation Method (BPM)}

The Beam Propagation Method solves the scalar Helmholtz equation under the paraxial approximation. The propagation of the electric field $E(x, y, z)$ in the direction $z$ can be described by:

\begin{equation}
\frac{\partial E(x, y, z)}{\partial z} = \frac{i}{2k_0} \nabla^2_{\perp} E(x, y, z) + i k_0 n_1(x, y) E(x, y, z)
\end{equation}

where $k_0$ is the free-space wave number, $\nabla^2_{\perp}$ is the transverse Laplacian operator, and $n_1(x, y)$ is the refractive index change.

\subsection{Finite-Difference Time-Domain (FDTD)}

The FDTD method discretizes Maxwell's curl equations in both time and space. For electric and magnetic fields, $\mathbf{E}$ and $\mathbf{H}$, Maxwell's equations can be represented as:

\begin{align}
\frac{\partial \mathbf{E}}{\partial t} &= \frac{1}{\epsilon_0} \nabla \times \mathbf{H} - \frac{\sigma}{\epsilon_0} \mathbf{E} \\
\frac{\partial \mathbf{H}}{\partial t} &= -\frac{1}{\mu_0} \nabla \times \mathbf{E}
\end{align}

where $\epsilon_0$ and $\mu_0$ are the permittivity and permeability of free space, respectively, and $\sigma$ is the electrical conductivity of the medium.

\subsection{Eigenmode Expansion (EME)}

In Eigenmode Expansion, the electric field $E$ in a waveguide can be expanded as a sum of eigenmodes $\phi_n$:

\begin{equation}
E(x, y, z) = \sum_n a_n(z) \phi_n(x, y) e^{i \beta_n z}
\end{equation}

where $a_n(z)$ are the mode amplitudes, $\phi_n(x, y)$ are the transverse field profiles of the eigenmodes, and $\beta_n$ are their propagation constants.

These formulas represent the core mathematical principles behind each simulation method, providing a basis for accurately modeling the propagation and interaction of light within photonic circuits.

\section{Simulation Software and Libraries}

Optical simulation software and libraries play a critical role in designing and analysing photonic components and systems, enabling researchers and engineers to predict the behavior of light within designed structures before physical fabrication. These tools vary widely in their capabilities, user interfaces, and specialisations, ranging from fully commercial packages to open-source platforms.

\subsection{Open-Source Software}

\begin{itemize}
    \item \textbf{Meep (MIT Electromagnetic Equation Propagation)} - Meep is a free and open-source software package for simulating electromagnetic systems, including photonic devices. It uses finite-difference time-domain (FDTD) methods to solve Maxwell’s equations in 2D or 3D. Meep is highly regarded for its flexibility and extensive features in optical simulations.
    \item \textbf{MPB (MIT Photonic Bands)} - MPB computes photonic band structures using a plane-wave expansion method. It is particularly useful for studying the properties of periodic photonic structures and is open-source, allowing users to explore and modify the code as needed.
\end{itemize}

\subsection{Commercial Software}

\begin{itemize}
    \item \textbf{COMSOL Multiphysics}: It is a finite element analysis based commercial simulation software package for photonics device design. It has the capability to simulate the coupled phenomena, or Multiphysics simulations involving charge and heat transfer modelling. COMSOL also offers an extensive interface to MATLAB and its toolboxes for a large variety of programming, pre-processing and post-processing possibilities.
    \item \textbf{Photon Design}:  Photon Design's FullWAVE is a simulation tool for studying the propagation of light in a wide variety of photonic structures, including integrated waveguide devices, as well as circuits and nanophotonic devices, such as photonic crystals. It combines methods based on semi-analytical techniques with other more numerical methods such as finite difference or finite element. FIMMWAVE comes with a range of user-friendly visual tools for designing waveguides, each optimized for a different geometry: rectangular geometries, circular geometries, and more general geometries e.g. diffused waveguides or other unusual structures.
    \item \textbf{CST MICROWAVE STUDIO (CST)}: It uses finite integration technique (FIT) to solve the Maxwell’s equations. A number of modules covering Transient, Frequency Domain, Integral Equation Solver, Heat transfer mechanical solver offering provide advantages in their own domains. 
    \item \textbf{RSoft}: RSoft product family includes: Component Design Suite to analyse complex photonic devices and components through industry-leading computer-aided design. It offers system Simulation to determine the performance of optical telecom and datacom links through comprehensive simulation techniques and component models.
    \item \textbf{Ansys Lumerical}: Lumerical offers a broad range of solver products (FDTD Solutions, MODE Solutions, , DEVICE, and Interconnect) that cover all physical simulation needs for integrated photonics.

\end{itemize}

\section{Quantum Computing}

\subsection{Quantum Software}

In the realm of quantum computing, which heavily utilizes photonics for quantum bit manipulation and other quantum operations, specific libraries have been developed:

\begin{itemize}
    \item \textbf{Pennylane (Xanadu)} - Pennylane is an open-source library developed by Xanadu for quantum machine learning, automatic differentiation, and optimisation of quantum circuits. It integrates seamlessly with existing machine learning libraries and is designed to enable quantum computing to be accessible in a classical computing framework.
    \item \textbf{Perceval (Quandela)} - Perceval is Quandela’s answer is tailored for the specific needs of photonic quantum computing, focusing on the nuances of single-photon sources and their applications in quantum technologies.
\end{itemize}
\label{secA1}

\end{appendices}





\end{document}